\documentclass{jpsj3}
\usepackage{txfonts}

\title{Superfluid, Supersolid, and Checkerboard Solid in Two-Component Bosons in 
an Optical Lattice: Study by Means of
Gross-Pitaevskii Theory and Monte - Carlo Simulations}

\author{Yoshihito Kuno, Keita Suzuki, and Ikuo Ichinose}
\inst{Department of Applied Physics, Nagoya Institute of Technology,
Nagoya 466-8555, Japan} 

\abst{The bosonic t-J model is a strong-on-site repulsion limit of the two-component
Bose-Hubbard model and is expected to be realized by experiments 
of cold atoms in an optical lattice.
In previous papers, we studied the bosonic t-J model by both 
analytical methods and numerical Monte - Carlo (MC) simulations.
However, in the case of finite $J_z$, where $J_z$ is the $z$-component coupling
constant of the pseudospin interaction, the phase diagram of the model was 
investigated by assuming the checkerboard type of boson densities.
In this study, we shall continue our previous study 
of the bosonic t-J model using both the Gross-Pitaevskii (GP) theory and 
MC simulations without assuming any pattern of boson densities.
These two methods complement each other and give reliable results.
We show that as $J_z$ is increased, the superfluid state 
evolves into a supersolid (SS), and furthermore 
into a genuine solid with the checkerboard symmetry.
In the present study, we propose a method identifying quantum phase transitions 
in the GP theory.
We also study finite-temperature phase transitions of the superfluidity
and the diagonal solid order of the SS by MC simulations.}

\begin{document}
\maketitle


\newcommand{\be}{\begin{eqnarray}}
\newcommand{\ee}{\end{eqnarray}}
\newcommand{\nn}{\nonumber\\}
\newcommand{\nin}{\noindent}
\newcommand{\la}{\langle}
\newcommand{\ra}{\rangle}

\section{Introduction}

Cold atomic gases in an optical lattice (OL) are one of the most actively 
studied systems these days because of their versatility\cite{review1}.
Sometimes they are regarded as a ``quantum simulator", e.g., for
the strongly-correlated many-body systems as they are highly 
controllable\cite{review2}.
The dimension and type of OL are controlled by adjusting the experimental
apparatus, the interactions between atoms are freely controlled by the Feshbach
resonance and using dipolar bosons etc\cite{dipole}.

It is now widely accepted that the physical properties of the system of bosonic
cold atomic gases in OL are well-described by the Bose-Hubbard model\cite{BHM}.
From the viewpoint of the strongly-correlated electron systems such as
the high-$T_C$ superconducting (SC) materials, the two-component Bose-Hubbard
model with the strong on-site repulsion is one of the most interesting systems.
Closely related systems can be realized by experiments on boson gases, 
e.g., $^{85}$Rb -$^{87}$Rb and $^{87}$Rb -$^{41}$K mixtures\cite{Rb2,RbK}.
As in the cuprates, it is expected that a pseudospin long-range order (LRO)
appears at low temperature.
Besides that magnetic order, a superfluid (SF) is also expected to form
at finite hole doping.
In general, the competition between the (antiferromagnetic) magnetic order and 
the superfluidity takes place, and then the system exhibits an interesting 
phase diagram.

In our previous papers\cite{KKI,KSI}, we showed that the strong on-site 
repulsion limit of 
the Bose-Hubbard model is described by the bosonic t-J model\cite{BtJ}.
We also discussed there that the bosonic t-J model appears from models
similar to  a bosonic counterpart of the d-p model for the strongly-correlated 
electron systems.
It is expected that a bosonic counterpart of the electronic d-p system
will be realized in the near future by experiments on Bose gas systems in the 
two-dimensional Lieb lattice.
It should be remarked that, in the bosonic t-J model as an effective low-energy
model of the bosonic d-p model, the parameters $t$ and $J$ are controlled almost freely
in contrast to the case of the cuprates. 

Using both analytical and numerical methods, we investigated the
phase diagram of the bosonic t-J model in the previous 
studies\cite{KKI,KSI}.
For numerical Monte-Carlo (MC) simulations, we assumed a homogeneous 
or checkerboard (CB) configuration for boson densities.
Amplitude modes of the bosons are then integrated out analytically,
and a quantum MC simulation was performed for the resultant effective
model for the phase degrees of freedom of the boson fields.
Then, we obtained the phase diagram of the bosonic t-J model.

In the present study, we shall continue the above studies and investigate
the phase diagram of the bosonic t-J model {\em without assuming any
density pattern of the bosons}.
We employ both the Gross-Pitaevskii (GP) theory and 
quantum MC simulations, in which the densities of bosons are treated as
dynamical variables.
The both methods are used to study the groundstate properties of the system
with Bose-Einstein condensation (BEC) and complement with each other.
In the present paper, we propose a method of identifying phase transitions
in the GP theory, which is found with the help of the MC simulation.

This paper is organized as follows.
In Sec.2, we introduce an extended version of the bosonic t-J model.
The GP equations and MC simulation for the bosonic t-J model are explained
in detail.
Section 3 is the main body of the present paper.
Numerical results of the GP equations and MC simulation are given.
We shall focus on the phase in which both bosons form BEC,
i.e., the 2SF state, and study how this state evolves as the antiferromagnetic (AF)
coupling $J_z$ is increased. 
In the first subsection, results of the GP equations are shown.
The phase transition from the 2SF to a supersolid (SS) is observed at some
critical value of the AF coupling by studying density snapshots and correlation functions. 
As the AF coupling is increased further, the transition from the SS to
a genuine solid without the SF long-range order (LRO) takes place.
In the second  and third subsections, results of the MC simulation are given.
Calculations of the internal energy and its quantum fluctuation,
which are in good agreement with those of the GP equations,
are shown, and the properties of observed phase transitions are discussed.
In the fourth subsection, finite-temperature ($T$) phase structures of the
SS and CB solid are investigated by the MC simulations.
As $T$ increases, the CB solid order disappears first
and then the long-range SF order does at a higher $T$.
Section 4 is devoted to the conclusion.

\section{Models: Bosonic t-J model and GP theory for two-component atoms}

As we explained in the introduction, we shall study the extended t-J model
on a square lattice whose Hamiltonian is given as
\begin{eqnarray}
&&H_{\rm EtJ}=H_{\rm tJ}+H_V, \label{HEtJ} \\
&&H_{\rm tJ}=-\sum_{\langle i,j\rangle} (t_a a^\dagger_{i}a_j
+t_b b^\dagger_{i}b_j+\mbox{h.c.})  
-J_{xy}\sum_{\langle i,j\rangle}(S^x_{i}S^x_j+S^y_{i}S^y_j)
+J_z\sum_{\langle i,j\rangle}S^z_iS^z_j  , \label{HtJ}  \\
&&H_V={V_0 \over 4}\sum_i\Big((a_i^\dagger a_i-\rho_{ai})^2+
(b_i^\dagger b_i-\rho_{bi})^2\Big),
\label{HV}
\end{eqnarray}
where $a^\dagger_i$ and $b^\dagger_i$ are 
boson creation operators at site $i$ of the square lattice and $t_a$ and $t_b$
are the hopping parameters.
The pseudospin operator $\vec{S}_i$ is given as  
$\vec{S}_i={1 \over 2}B^\dagger_i\vec{\sigma}B_i$ with
$B_i=(a_i,b_i)^t$, and $\vec{\sigma}$ is the Pauli spin matrix.
In the t-J model, the doubly-occupied state is excluded at each site
as it is derived from the Bose-Hubbard model in the strong on-site
repulsion limit.
We add the on-site repulsive terms $H_V$ of the Hubbard type, which control fluctuations
in the number of the particles at each site, although it is expected that this term
substantially appears from the effects of the terms in $H_{\rm tJ}$ in 
Eq.(\ref{HtJ}), particularly the $J$-terms.
For more details,
see later discussion on the relationship between the extended t-J model and the
GP theory.
Then, in $H_V$, $V_0$ is a positive parameter and $\rho_{ai}+\rho_{bi}\leq 1$.
To impose the local constraint 
$a^\dagger_ia_i+b^\dagger_ib_i<1$ on the physical state in the Hilbert space, 
the {\em slave-particle representation} is useful.

In our previous paper\cite{KKI}, we studied the phase diagram of the model
$H_{\rm EtJ}$ by the MC simulations in the slave-particle
representation.
The case of $J_z\ll 1$ can be studied straightforwardly 
as the integration over the amplitude modes of the bosons $a_i$ and 
$b_i$ can be carried out without any difficulties, and the action of the resultant 
model of the phase degrees of freedom is positive-definite.
We call the resultant model the extended XY model.

On the other hand, the case of finite $J_z$ has been investigated in Ref.8.
In that study, we focused on the appearance of the CB-type density pattern 
of the bosons and found that a SS state forms for intermediate values of
$J_z$ although for a large $J_z$ the genuine CB solid without the SF appears
as the groundstate.
In that study, we assumed the density pattern such as 
$\rho_{a\ {\rm even-site}}=\rho_{b\ {\rm odd-site}}$ and 
$\rho_{a\ {\rm odd-site}}=\rho_{b\ {\rm even-site}}$,
and therefore the considered states have the translational symmetry of 
the twofold lattice spacing. 

In the present research, we also study a GP theory for a system of 
two-component bosons 
trapped in an OL, which is regarded as a genuine system in experiments
for the t-J model on the lattice.
In that study, we do not assume any translational symmetry for solutions.
It will become clear that the results of the GP equations
are helpful for performing reliable MC simulations for the t-J model.

\begin{figure}[h]
\begin{center}
\includegraphics[width=6cm]{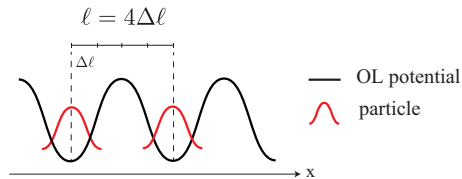}
\vspace{-0.3cm}
\caption{(Color online)
Optical lattice described by the potential $V_{\rm OL}$ in Eq.(\ref{VOL}).
Long-range interactions between atoms in adjacent minima of OL
are given by Eq.(\ref{intOL}).
$\Delta \ell$ is the spatial slice used for solving the GP equations.
}\vspace{-0.5cm}
\label{fig:LRint}
\end{center}
\end{figure}

We use the continuum description  of the system and introduce the following
periodic potential in the GP equations to simulate the square OL, i.e., 
\begin{equation}
V_{\rm OL}=A_{\rm OL}\Big(\sin^2(\pi x/\ell)+\sin^2(\pi y/\ell)\Big),
\label{VOL}
\end{equation}
where the positive parameter $A_{\rm OL}$ is the depth of the OL and
$\ell$ is the lattice spacing, which are determined by the strength and wavelength of 
the laser used in experiments, respectively.
See Fig.\ref{fig:LRint}.
By using the OL potential Eq.(\ref{VOL}), Hamiltonian of the $a$ and $b$ atoms 
is given as
\begin{eqnarray}
H_{\rm GP}&=&\int d^2x\Big[
\sum_{\alpha=a,b}\psi^\dagger_\alpha\Big(-{\hbar^2 \over 2m}
\nabla^2+V_{\rm OL}\Big)\psi_\alpha 
-g_{aa}|\psi_a|^2(1-|\psi_a|^2)-g_{bb}|\psi_b|^2(1-|\psi_b|^2) \nonumber \\
&&+g_{ab}|\psi_a|^2|\psi_b|^2\Big],
\label{HGP1}
\end{eqnarray}
where $g_{aa}$'s are the intra- and inter-repulsive coupling constants.
In the present study, we mostly consider the case $g_{aa}=g_{bb}>g_{ab}$
in which two atoms are miscible.

The Hamiltonian $H_{\rm GP}$ in Eq.(\ref{HGP1}) describes the system of 
two-component BEC in the OL and
with the intra- and inter-species on-site repulsive interactions.
Then, we add the following long-range interactions between atoms 
to the Hamiltonian of the GP theory, which
corresponds to the $J$-terms in the t-J model $H_{\rm tJ}$,
\begin{eqnarray}
&&H_{\rm LR}=\int d^2x {\cal H}_{\rm LR}, \nonumber \\
&&{\cal H}_{\rm LR}=   
\sum_{|x-y|=\ell}\Big[
-j_{xy}\psi^\dagger_a(x)\psi_b(x)\psi_a(y)\psi^\dagger_b(y)  
+\mbox{H.c.} \label{intOL} \\
&&\hspace{0.5cm}+j_z(|\psi_a(x)|^2-|\psi_b(x)|^2)
(|\psi_a(y)|^2-|\psi_b(y)|^2)\Big],
\nonumber
\end{eqnarray}
where $j_{xy}$ and $j_z$ are parameters corresponding to $J_{xy}$ and $J_z$,
respectively, and
$|x-y|=\ell$ denotes that the distance between sites
$x$ and $y$ is $\ell$.
See Fig.\ref{fig:LRint}.
Then, the total Hamiltonian $H_{\rm GP}+H_{\rm LR}$ gives the following 
GP equations\cite{GPE}:
\begin{eqnarray}
(i-\gamma)\hbar {\partial \psi_a\over \partial t} &=& 
\Big[-{\hbar^2 \over 2m}\nabla^2+V_{\rm OL}+g_{aa}|\psi_a|^2
+g_{ab}|\psi_b|^2-\mu\Big]\psi_a \nonumber \\
&&+\sum_{y, |x-y|=\ell}
\Big[-j_{xy}\psi^\ast_b(x)\psi_b(y)\psi_a(y) 
+j_z\psi_a(x)(|\psi_a(y)|^2-|\psi_b(y)|^2)\Big], \nonumber\\
 (i-\gamma)\hbar {\partial \psi_b\over \partial t} &=&  
\Big[-{\hbar^2 \over 2m}\nabla^2+V_{\rm OL}+g_{bb}|\psi_b|^2
+g_{ba}|\psi_a|^2-\mu\Big]\psi_b  \nonumber \\
&&+\sum_{y, |x-y|=\ell}
\Big[-j_{xy}\psi^\ast_a(x)\psi_a(y)\psi_b(y) 
+j_z\psi_b(x)(|\psi_b(y)|^2-|\psi_a(y)|^2)\Big],
\label{GPEq2}
\end{eqnarray}
where $\mu$ is the chemical potential and $\gamma$ is a phenomenological 
dissipative damping parameter.
In the practical calculation, we set $\gamma=0.1$, for which the convergence of solutions 
is obtained most smoothly in the time evolution of the GP equations, although 
its value does not substantially affect the physical results.
We think that 
this result comes from the fact that {\em we are studying stationary states in a
system with stable low-energy states}. 
For the study of a system in an external magnetic field (an effective magnetic
field can be generated by rotating boson gas systems), which was discussed in 
Ref.10 etc, a smaller $\gamma$ has to be used to
obtain reliable results, as the system exhibits nontrivial dynamical behaviors  
by the appearance of vortices etc.

In a practical numerical calculation,
we first turn off $H_{\rm LR}$ to verify that bosons tend to locate in 
the minima of the OL, and then turn on $H_{\rm LR}$ to see the 
effects of the $j$-terms.
The differential equations in Eq.(\ref{GPEq2}) are converted to 
difference equations as
${\partial \psi \over \partial x}\Rightarrow 
(\psi(x+\Delta \ell)-\psi(x))/\Delta \ell$, where the 
spatial slice $\Delta\ell$ is taken as 
$\Delta\ell=\ell/4$ and is set to unity in the calculation, $\Delta\ell=1$.
Then, $\ell=4$ in this unit.
See Fig.\ref{fig:LRint}.
Similarly, the time slice $\Delta t$ is used as
${\partial \psi \over \partial t}\Rightarrow 
(\psi(t+\Delta t)-\psi(t))/\Delta t$ and is set to $\Delta t=10^{-4}$ in the 
numerical study.
Natural unit of time is ${2m(\Delta\ell)^2 \over \hbar}$ and
it is estimated as ${2m(\Delta\ell)^2 \over \hbar}\sim (10^{-5}-10^{-4})$ s
for typical experiments.
From this observation, we used the value $\Delta t=10^{-4}$.
We have verified the stability of the obtained solutions of the GP equations
as varying values of $\Delta t$ and $\gamma$, such as 
$\Delta t=10^{-5}$ and $\gamma=0.05$.

It is instructive to compare the above GP equations to those 
of the t-J model derived from Eqs.(\ref{HEtJ}), (\ref{HtJ}) and (\ref{HV}),
\begin{eqnarray}
(i-\gamma)\hbar{\partial a_i \over \partial t} &=&
-t_a\sum_{j\in iNN}a_j-2J_{xy}\sum_{j\in iNN}b_ia_jb^\dagger_j
+J_z\sum_{j\in iNN}a_i(a^\dagger_ja_j-b^\dagger_jb_j) \nonumber \\
&&+{V_0 \over 2}a_i(a^\dagger_ia_i-\rho_{ai}), \nonumber
\end{eqnarray}
\begin{eqnarray}
(i-\gamma)\hbar{\partial b_i \over \partial t} &=&
-t_b\sum_{j\in iNN}b_j-2J_{xy}\sum_{j\in iNN}a_ib_ja^\dagger_j
-J_z\sum_{j\in iNN}b_i(a^\dagger_ja_j-b^\dagger_jb_j) \nonumber \\
&&+{V_0 \over 2}b_i(b^\dagger_ib_i-\rho_{bi}),
\label{GPtJ}
\end{eqnarray}
where $j\in iNN$ stands for the nearest-neighbor (NN) sites of $i$.
We identify the lattice spacing of the OL $\ell\sim a_L$, where $a_L$ 
is the lattice spacing of the square lattice on which the t-J model
is defined.
As $[\psi]=$ (dimension of $\psi)=(\mbox{length})^{-1}$ and
$[a]=[b]=(\mbox{length})^{0}$, we have
$\psi_\alpha\sim \alpha/{\ell} \; (\alpha=a,b)$.
Then, the following straightforward correspondence between the parameters 
in the two systems is obtained:
\begin{equation}
j_{xy}\sim 2J_{xy}/\ell^2,  \;\; j_z \sim J_z/\ell^2, \;\; 
g_{aa}=g_{bb}\sim V_0/(2\ell^2).
\label{cores}
\end{equation}
However, a practical calculation shows that the nonlocal $j$-terms ($J$-terms)
strongly suppress the density fluctuations of atoms at each site. 
See for example, Fig.\ref{fig:GP0}.
Therefore, $H_V$ in the extended t-J model acquires the renormalization effect
from the $J$-terms and can be regarded as the leading term.

Finally, let us briefly explain the MC simulation of the t-J model described by the 
slave-particle representation\cite{KKI},
\begin{eqnarray}
&& a_i=\phi^\dagger_i \varphi_{i1}, \;\;\; 
b_i=\phi^\dagger_i \varphi_{i2},  \label{slave}  \\
&& \Big(\phi^\dagger_i\phi_i+\varphi^\dagger_{i1}\varphi_{i1}+
\varphi^\dagger_{i2}\varphi_{i2}-1\Big)
|\mbox{Phys}\rangle =0,
\label{const}
\end{eqnarray}
where $\phi_i$ is a boson operator that {\em annihilates hole} at site $i$,
whereas $\varphi_{1i}$ and $\varphi_{2i}$ are bosons that represent the pseudospin 
degrees of freedom.
$|\mbox{Phys}\rangle$ is the physical state of the slave-particle Hilbert space.
Then, the partition function is given by the path-integral formalism as
\begin{eqnarray}
Z&=& \int [D\phi D\varphi_1 D\varphi_2]
\exp\Big[-\int d\tau\Big(\bar{\varphi}_{1i}(\tau)\partial_\tau \varphi_{1i}(\tau)
\nonumber \\
&&+\bar{\varphi}_{2i}(\tau)\partial_\tau \varphi_{2i}(\tau)
 +\bar{\phi}_i(\tau)\partial_\tau \phi_i(\tau)  
+H_{\rm EtJ}
\Big)\Big],
\label{Z}
\end{eqnarray}
where $\tau$ is the imaginary time, 
$H_{\rm EtJ}$ is expressed by the slave particles, and 
the above path integral is calculated under the constraint given by Eq.(\ref{const}).
Before performing the MC simulation, we parameterize
$\varphi$'s and $\phi$ as ($\sum_{a=1,2,3}\rho_{ai}=1$)
\begin{eqnarray}
\varphi_{1i}=\sqrt{\rho_{1i}+\ell_{1i}}\exp(i\omega_{1i}), \; \; 
\varphi_{2i}=\sqrt{\rho_{2i}+\ell_{2i}}\exp(i\omega_{2i}), \; \;
\phi_i=\sqrt{\rho_{3i}+\ell_{3i}}\exp(i\omega_{3i}), 
\label{param1}
\end{eqnarray}
and integrate out the radial degrees of freedom controlled by the term $H_V$.
The constraint $\ell_{1i}+\ell_{2i}+\ell_{3i}=0$
can be incorporated by using a Lagrange multiplier $\lambda_i(\tau)$.
The variables $\ell_{\sigma i} \ (\sigma=1,2,3)$ also appear in $H_{\rm tJ}$,
but we ignore them by simply replacing 
$\varphi_{\sigma i} \rightarrow \sqrt{\rho_{\sigma i}}\exp(i\omega_{\sigma i})$, 
and then we have
\begin{eqnarray}
\int d\lambda_i d\ell_{i}e^{\int d\tau\sum_{\sigma=1}^3(-V_0(\ell_{\sigma,i})^2
+i\ell_{\sigma,i}(\partial_\tau \omega_{\sigma,i}+\lambda_i))}
=\int d\lambda_i e^{-{1 \over 4V_0}\int d\tau\sum_\sigma
(\partial_\tau \omega_{\sigma,i}+\lambda_i)^2}.
\label{integral}
\end{eqnarray}
The resultant quantity on the RHS of Eq.(\ref{integral}) is positive-definite,
and therefore the numerical study by the MC simulation can be carried out
without any difficulty.
It should be remarked that the Lagrange multiplier $\lambda_i$
in Eq.(\ref{integral}) behaves as a gauge field, i.e.,
the RHS of Eq.(\ref{integral}) is invariant under the following ``gauge transformation",
$\omega_{\sigma,i}\rightarrow \omega_{\sigma,i}+\alpha_i, \ 
\lambda_i \rightarrow \lambda_i-\partial_\tau \alpha_i$.

The partition function $Z$ in Eq.(\ref{Z}) depends on the local density 
of the bosons $\rho_{ai}$.
In the previous studies, we assumed a homogeneous distribution of bosons
in the case of $J_z\ll 1$ or the CB symmetry for a finite $J_z$
and treated the {\em global density difference}
$\Delta\rho\equiv \rho_{1{\rm even-site}}-\rho_{1{\rm odd-site}}
=-(\rho_{2{\rm even-site}}-\rho_{2{\rm odd-site}})$ as a
variational parameter.
These treatments obviously preclude the possibility of, for example., a phase-separated 
state.
In the present study, we shall treat the {\em local densities} $\rho_{ai}$ 
as variation variables
and determine them using the requirement of the minimum free-energy condition.
This means that not only the $V$-term in $H_V$ but also the $t$ and $J$-terms
affect the local density of the bosons.
Numerical studies in Sec.3 show that the homogeneous or
the CB configuration of the boson density dominates in most of 
the parameter regions of the models as we assumed in our previous study\cite{KSI}.

\section{Numerical results: Solutions to GP theory and MC simulation of t-J model}

In this section, we shall show the results obtained by the numerical calculations.
We first show numerical solutions of the GP equations.
We start with the state with the BECs of both the $a$ and $b$ atoms
and then we increase $j_z$ and see how the state evolves.  
Solutions of the GP equation show that a SS appears at some critical
value of $j_z$.
As $j_z$ is increased further, the BECs disappear, and
the genuine solid with only the CB order replaces the SS.
In the second and third subsections of this section, we show the results of the 
MC simulations.
We first consider the case of an equal-mass $t_a=t_b=t$, and show that
the results are in good agreement with those of the GP equations.
Calculations of the correlation functions by the MC simulations identify unambiguously
existing orders in each phase.
Then, we apply a similar MC analysis in the case of a different mass such as $t_a=2t_b$,
and clarify the phase diagram.
Finally, in the last subsection, we study the finite-temperature phase
diagram of the SS and CB solid and see how two orders, i.e., the SF and CB, disappear
by the thermal fluctuations as the temperature increases.
All the calculations were carried out for the system at a filling factor $0.35$ 
for each atom (therefore, the total filling factor is $0.7$).

\subsection{Gross-Pitaevskii theory}

\begin{figure}[h]
\begin{center}
\includegraphics[width=5cm]{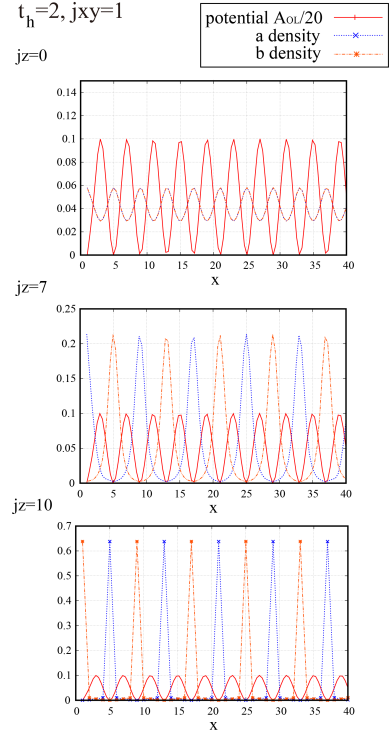}
\vspace{-0.3cm}
\caption{(Color online)
Density profiles of solutions of GP equations. 
Parameters are $t_h\equiv {\hbar^2\over m}=2, g_{aa}=g_{bb}=0.5, 
g_{ab}=0$, and $j_{xy}=1$.
For $j_z=0$, both $a$ and $b$-atoms are located in each site of the OL, whereas
the CB pattern forms for $j_z=10$. 
$\ell$ is the lattice spacing of the OL and $\ell=4$ in the numerical calculation.
}\vspace{-0.5cm}
\label{fig:GP0}
\end{center}
\end{figure}

\begin{figure}[h]
\begin{center}
\includegraphics[width=8cm]{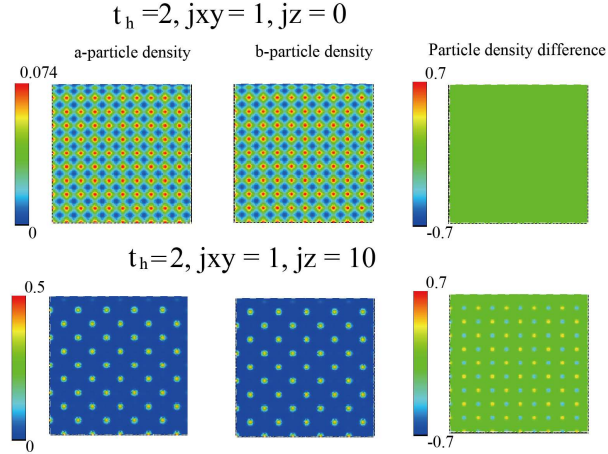}
\vspace{-0.3cm}
\caption{(Color online)
Snapshots of solutions of GP equations, where
$a$-particle density per $(\Delta \ell\times \Delta \ell)$ square$=|\psi_a(x)|^2$, etc.
Hopping parameter $t_h\equiv {\hbar^2\over m}$.
For $j_z=0$, the state is homogeneous, whereas the CB solid forms for $j_z=10$.
}\vspace{-0.5cm}
\label{fig:GP1}
\end{center}
\end{figure}

In this subsection, we shall show solutions of the GP equations in Eq.(\ref{GPEq2}).
We started with the case of $j_z=0$ and increased $j_z$ gradually\cite{initial}.
The depth of the OL was set to $A_{\rm OL}=2$.
For $j_z=0$, all the bosons are trapped in the OL, and the homogeneous state 
with the double BECs forms for a sufficiently large 
$t_h\equiv {\hbar^2\over m(\Delta \ell)^2}={\hbar^2\over m}$.
See the density profiles 
$|\psi_a(x)|$ and $|\psi_b(x)|$ in Figs.\ref{fig:GP0} and
\ref{fig:GP1}.
This state corresponds to the 2SF state with the ferromagnetic (FM) 
pseudo-spin order for $j_{xy}>0$,
which was observed previously in the MC simulations\cite{KSI}.
By the practical calculation, we verified that for every configuration (except
the phase-separated state, see later discussion),
a local constraint similar to that of the t-J model is satisfied as
\begin{equation}
\int_{x\in {\rm site \ of \ OL}} d^2x (|\psi_a(x)|^2+|\psi_b(x)|^2)\simeq 0.7,
\label{GPlocalconst}
\end{equation}
where $\int_{x\in {\rm site \ of \ OL}}$ denotes that the integral over 
the $(\ell\times \ell)$ region of a single site of the OL.
See Fig.\ref{fig:GP0}.
The average hole density is, therefore, $30\%$ as we explained above.
The stability of Eq.(\ref{GPlocalconst}) is guaranteed by the conservation of the
total number of particles in the time evolution of the GP equations.


\begin{figure}[t]
\begin{center}
\includegraphics[width=7.5cm]{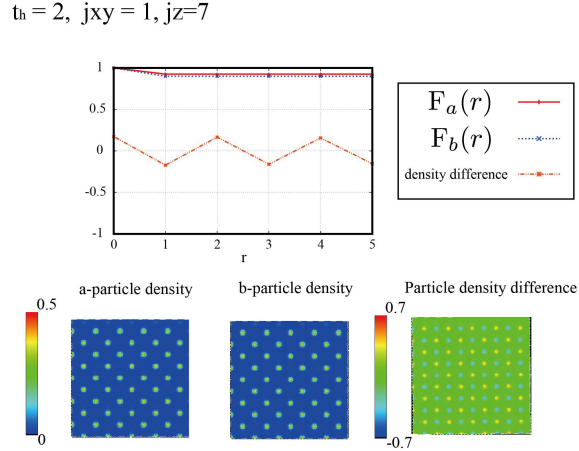}
\vspace{-0.3cm}
\caption{(Color online)
Correlation functions of solutions of GP equations and
particle density per $(\Delta \ell\times \Delta \ell)$ square.
Both the SF and CB orders exist, i.e., the SS forms for $j_z=7.0$.
}\vspace{-0.5cm}
\label{fig:GP2}
\end{center}
\end{figure}

We started with the above 2SF state and increased $j_z$
to see how the state evolves.
We found that the SS appears at a certain critical value $j_z$, $j_{z1}\simeq 6.0$. 
There the density difference $|\psi_a(x)|^2-|\psi_b(x)|^2\neq 0$ appears, 
but the phase coherence of the $a$ and $b$-atoms still exists, as verified
by calculating the following correlation functions:
\begin{equation}
F_\alpha(r)=\Big\langle {1 \over 2|\psi_\alpha(x)\psi_\alpha(x+r)|}
(\psi^\ast_\alpha(x)\psi_\alpha(x+r)+\mbox{c.c.})\Big\rangle,
\label{F}
\end{equation}
where $\alpha=a, b$, and $x$ and $x+r$ are both sites of the OL.
$F_\alpha(r)$ in Eq.(\ref{F}) was evaluated by substituting the obtained ``wave
functions" $\{\psi_a(x), \psi_b(x)\}$ after a sufficiently long evolution period
in the GP equations.
For $j_z=7$, see Fig.\ref{fig:GP2}.

We increased $j_z$ further.
At the second critical value of $j_z$, $j_{z2}\simeq 8.0$, 
the superfluidity of the atoms is lost and the genuine solid with the
CB pattern appears instead of the SS.
In Figs.\ref{fig:GP1} and \ref{fig:GP2}, 
we show snapshots of each phase and also correlation functions
in the SS.
We also verified that, for a much larger $j_z\gg A_{\rm OL}$ and $1/m$, 
the phase separation into the region of the pure CB solid and the hole-rich region 
takes place, i.e., {\em Eq.(\ref{GPlocalconst}) is not satisfied in that state}.

The above result indicates that {\em quantum phase transitions} take
place as $j_z$ is increased.
It is interesting and also challenging to explore how the quantum phase
transition is identified from solutions of the GP equations.
As the MC simulation can identify the quantum phase transition by calculating
the expectation value of energy and its quantum fluctuations,
we propose to observe similar quantities in solutions of the GP equations.
The energy of the state obtained by the GP equations, $E_{\rm GP}$,
is naturally defined as
\begin{eqnarray}
&&E_{\rm GP}=\langle H^\ast_{\rm GP} \rangle/N_s, \nonumber \\
&& H^\ast_{\rm GP}\equiv 
H_{\rm GP}+H_{\rm LR}-
\int d^2x \sum_{\alpha=a,b}\psi^\dagger_\alpha V_{\rm OL}\psi_\alpha,
\label{EGP}
\end{eqnarray}
where $N_s$ is the number of sites of OL and 
the expectation value is evaluated using the boson wave function that is obtained after
a fairly long-time evolution of the GP equations.
In the definition of $E_{\rm GP}$, we subtract the energy of the OL,
and therefore  $E_{\rm GP}$ corresponds to $\langle H_{\rm EtJ} \rangle$
in the MC simulation\cite{FN_EGP}. 

\begin{figure}[h]
\begin{center}
\includegraphics[width=5cm]{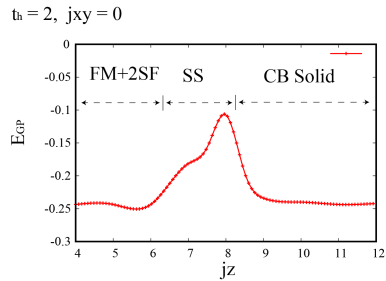}
\vspace{-0.3cm}
\caption{(Color online)
Energy as a function of $j_z$ calculated using the GP equations.
The increase in $E$ in the SS state comes from the fact that the hopping term 
and the $j_z$-term in the Hamiltonian $H_{\rm GP}+H_{\rm LR}$ compete 
with each other.
}\vspace{-0.5cm}
\label{fig:GP3}
\end{center}
\end{figure}

In Fig.\ref{fig:GP3}, the obtained result of $E_{\rm GP}$ is shown as a function of $j_z$,
and this result will be compared with that obtained by the quantum
MC simulation in the subsequent subsection.
The behavior of $E_{\rm GP}$ in Fig.\ref{fig:GP3} indicates the existence of two 
phase transitions at $j_z\simeq 6.0$ and $8.0$.
We verified that a behavior of $E_{\rm GP}$ similar to
that shown in Fig.\ref{fig:GP3} is obtained for the solution $\{\psi_a(x), \psi_b(x)\}$
to the GP equation at various times $t$.
We carefully investigated the GP solutions at various $j_z$ values, 
and verified that, at these two $j_z$'s, phase transitions from the SF to the SS, 
and also from the SS to the CB solid take place.

\begin{figure}[h]
\begin{center}
\includegraphics[width=5cm]{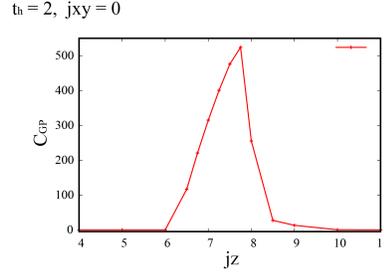}
\vspace{-0.3cm}
\caption{(Color online)
Specific heat as a function of $j_z$ calculated using the GP equations.
}\vspace{-0.5cm}
\label{fig:GP4}
\end{center}
\end{figure}

It is interesting to see how the fluctuation of the energy behaves at the above
values of $j_z$.
However, the straightforward definition of the specific heat $C$ such as 
$C=(\langle (H^\ast_{\rm GP})^2 \rangle-\langle H^\ast_{\rm GP}\rangle^2)
/N_s$ {\em always vanishes} when it is evaluated using the wave function 
$\{\psi_a(x), \psi_b(x)\}$.
Then, we define ``specific heat" $C_{\rm GP}$ of the present system 
from the viewpoint of the path-integral formulation of the system.
In the path integral, the partition function of  system with the Hamiltonian $H$
is given as
\begin{equation}
Z_H=\int [d\phi] \ e^{\int d\tau A_c},  \;\; 
\langle H \rangle=\int [d\phi] \ He^{\int d\tau A_c}/Z,
\label{general}
\end{equation}
where action $A_c=(\mbox{Berry phase})-H$.
Let us assume that there exists a coupling constant $g$ in $H$, and a phase transition 
takes place as $g$ is varied.
From Eq.(\ref{general}), the ``specific heat" $C_g$ is defined as
\begin{eqnarray}
C_g&=&
{\partial \over \partial g}\langle H \rangle  \nonumber  \\
&=&
\Big\langle {\partial H \over \partial g}\Big\rangle
-\Big\langle H\int d\tau \ {\partial H\over \partial g}\Big\rangle
+\langle H\rangle\Big\langle\int d\tau \ {\partial H \over \partial g}\Big\rangle.
\label{Cg}
\end{eqnarray}
Assuming a short-range correlation, 
\begin{equation}
\Big\langle H\int d\tau \ {\partial H\over \partial g}\Big\rangle
-\langle H\rangle\Big\langle\int d\tau \ {\partial H \over \partial g}\Big\rangle
\propto O(N_s),
\label{Cg2}
\end{equation}
as the genuine specific heat 
${\partial\langle H \rangle \over \partial T}
\propto \langle H^2\rangle-\langle H\rangle^2\propto O(N_s)$.
From the above consideration, we propose to use the following $C_{\rm GP}$
for observing phase transitions:
\begin{equation}
C_{\rm GP}=\Big\langle \Big({1 \over N_s}{\partial H^\ast_{\rm GP} \over \partial j_z}
-{1 \over N^2_s}H^\ast_{\rm GP}{\partial H^\ast_{\rm GP} \over \partial j_z}\Big)
\Big\rangle. 
\label{CGP}
\end{equation}
The calculation of $C_{\rm GP}$ is shown in Fig.\ref{fig:GP4}.
We again verified that the same behavior of $C_{\rm GP}$ to
that shown in Fig.\ref{fig:GP4} is obtained for the solution $\{\psi_a(x), \psi_b(x)\}$
to the GP equation at different times $t$.
$C_{\rm GP}$ has a peak at $j_z\simeq 8.0$ as $E_{\rm GP}$ does.
This result seems to indicate that the phase transition from the SS to
the CB solid is of second order, whereas that from the 2SF to the SS
is a crossover or higher-order phase transition as the derivative of $C_{\rm GP}$ 
with respect to $j_z$ shows an anomalous behavior at $j_z\simeq 6.0$.
This observation will be confirmed by the MC simulations explained in the 
subsequent subsection.


\subsection{Monte-Carlo simulations: Case of the same mass}

In this subsection, we shall study the extended bosonic t-J model $H_{\rm EtJ}$
by the MC simulations.
We take $a_L$ as the unit of length in the lattice model and will often set $a_L=1$ in 
this and subsequent subsections.
(Here note that this definition of the unit of length is
different from that used in the GP theory, $\Delta \ell=1$.)
For the MC simulations, we also introduce a lattice for the imaginary time $\tau$
with the lattice spacing $\Delta\tau$.
Then, the model is defined on the three-dimensional (3D) space-time lattice,
and we denote the site of the 3D lattice $r$.
The study of the GP equations in the previous subsection strongly
suggests that holes are distributed homogeneously except for a very large
$J_z$ compared with the hopping parameters $t_a$ and $t_b$.
Therefore, for the practical numerical calculation, we assume a 
homogeneous distribution of holes and fix the hole density at each site to 
$30\%$, i.e., $\rho_{3r}=\rho_3=0.3$, as in the GP theory in the previous
subsection.
Then, we consider the model whose partition function $Z_{qXYZ}$ is given as
\begin{eqnarray}
Z_{qXYZ}&=&\int \prod_{a=1,2,3}[d\omega_{ar}][d\lambda_r]e^{A_{qXYZ}},
\nonumber \\
A_{qXYZ}&=&A_{\tau}+A_{\rm L}(e^{i\Omega_\sigma},e^{-i\Omega_\sigma})
+A_z,
\label{AqXYZ}
\end{eqnarray}
where
\begin{eqnarray}
A_{\tau}=-c_\tau\sum_{r} \sum_{\sigma=1}^3
\cos (\omega_{\sigma,r+\hat{\tau}}-\omega_{\sigma r}+\lambda_r),  
\label{Atau}
\end{eqnarray}
\begin{eqnarray}
A_{\rm L}(e^{i\Omega_\sigma},e^{-i\Omega_\sigma}) 
 &=&\sum_{\langle r,r'\rangle_S}
\Big(C_1\cos (\Omega_{1,r}-\Omega_{1,r'})
+C_2\cos (\Omega_{2,r}-\Omega_{2,r'})  \nonumber \\
&&+C_3\cos (\Omega_{3,r}-\Omega_{3,r'})\Big),
\label{AL2}
\end{eqnarray}
and
\begin{eqnarray}
A_z=-J_z\sum_{\langle r,r'\rangle_S}\Delta\rho_{r}\Delta\rho_{r'}, \;\; 
\Delta\rho_r\equiv\rho_{1,r}-\rho_{2,r},
\label{Az}
\end{eqnarray}
where $\langle r,r'\rangle_S$ denotes the NN sites in the 2D spatial lattice.
The dynamical variables $\Omega_{a, r}\ (a=1,2,3)$ are related to the phases
$\omega_{ar}$ as
$$
\Omega_{1,r}=\omega_{1 r}-\omega_{2 r}, \
\Omega_{2,r}=\omega_{1 r}-\omega_{3 r}, \
\Omega_{3,r}=\omega_{2 r}-\omega_{3 r}.
$$
The partition function in Eq.(\ref{AqXYZ}) has been derived by 
integrating out the amplitude
modes of the slave-particle fields as explained in Sec.2.
Then, the coefficients in the action $A_{qXYZ}$ depend on the density difference
$\Delta\rho_r$ and are given as
\begin{eqnarray}
&&c_\tau={1 \over V_0\Delta\tau},  \nonumber \\
&&C_1=J_{xy}\rho_3\Delta\tau\sqrt{((1-\rho_3)^2-(\Delta\rho_r)^2)
                     ((1-\rho_3)^2-(\Delta\rho_{r'}))}, \nonumber \\
&&C_2=t_a\rho_3\Delta\tau\sqrt{(1-\rho_3+\Delta\rho_r)
                             (1-\rho_3+\Delta\rho_{r'})},  \nonumber  \\
&&C_3=t_b\rho_3\Delta\tau\sqrt{(1-\rho_3-\Delta\rho_r)
                             (1-\rho_3-\Delta\rho_{r'})}.
\label{C123}
\end{eqnarray}               
From the relation $1/(k_{\rm B}T)=L\cdot\Delta\tau$, $\Delta\tau$ has 
dimension $1/$(energy) and the low-temperature limit
is realized for $L\rightarrow \infty$.
The parameters $c_\tau, \sim, C_3$ in Eq.(\ref{C123}) are dimensionless, and we put $c_\tau=2$ for the practical calculation. 
Then, $k_{\rm B}T=(c_\tau V_0)/L=2V_0/L$.

$Z_{qXYZ}$ in Eq.(\ref{AqXYZ}) is a functional of $\{\Delta\rho_r\}$,
i.e., $Z_{qXYZ}=Z_{qXYZ}(\{\Delta\rho_r\})$.
We expect that $\{\Delta\rho_r\}$ behave as variational variables and 
determine them under the optimal free-energy condition.
In the practical calculation, we performed the local update of $\{\Delta\rho_r\}$
by the MC simulation and obtained 
\begin{equation}
[Z_{qXYZ}]\equiv \int [d\Delta\rho_r]Z_{qXYZ}(\{\Delta\rho_r\}).
\label{ZqXYZ}
\end{equation}
However,
in the MC calculations, we found that {\em $\{\Delta\rho_r\}$ is quite stable} for 
given values of the parameters in the action $A_{qXYZ}$.
This fact indicates that $\{\Delta\rho_r\}$ should be regarded as variational
parameters rather than dynamical variables.

In the following, we shall show the results for $\rho_3=0.3$ as 
stated above.
For the MC simulations, we employ the grand-canonical ensemble,
and therefore, the total numbers of $a$ and $b$-atoms, $N_a$ and $N_b$,
are {\em not conserved separately} in each MC update, although $N_a+N_b$
{\em is conserved}.
We start with the state of the ferromagnetic (FM)+2SF for small 
$\tilde{J}_z\equiv J_z\Delta \tau$
and increase $\tilde{J}_z$ gradually and see how the phase evolves.
For numerical simulations, 
we employ the standard Monte-Carlo Metropolis algorithm
with local update\cite{Met}.
The typical sweeps for the measurement is $(30000 - 50000)\times (10$ samples),
and the acceptance ratio is $40 - 50\%$.
Errors are estimated from 10 samples by the jackknife methods\cite{jack}.

\begin{figure}[h]
\begin{center}
\includegraphics[width=5cm]{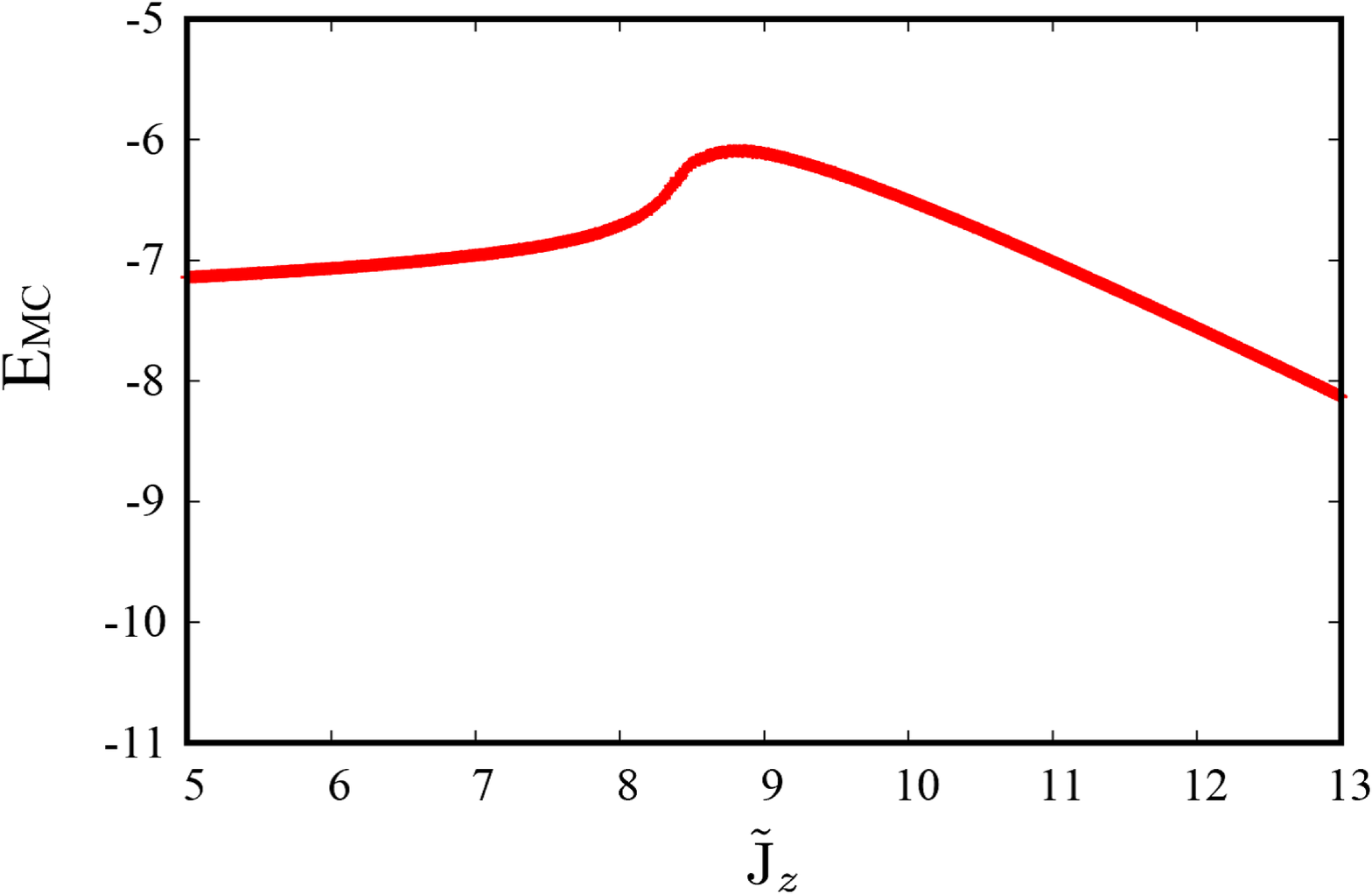}
\includegraphics[width=5cm]{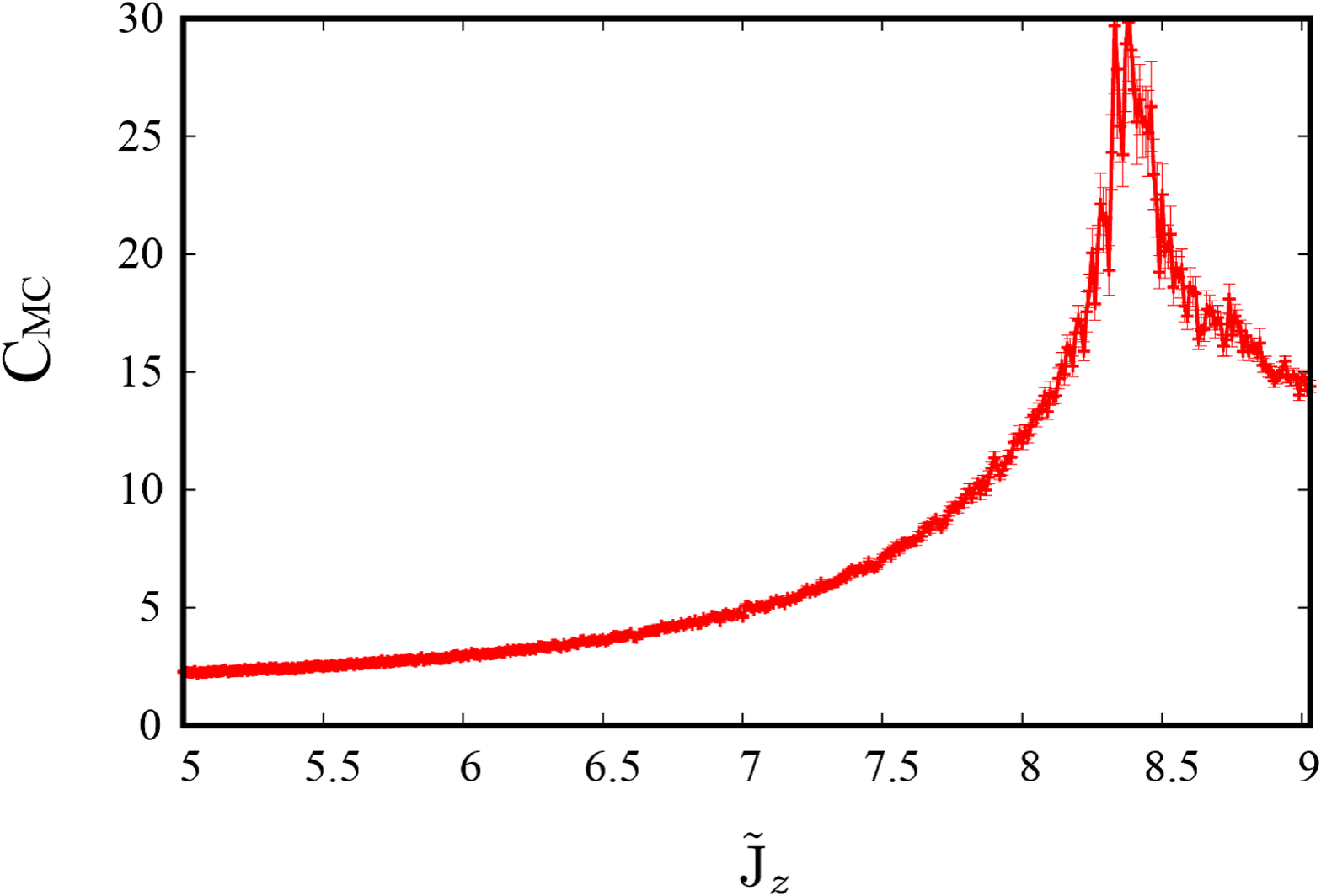}
\vspace{-0.3cm}
\caption{(Color online)
Internal energy and specific heat of qXYZ model as a
function of $\tilde{J}_z\equiv J_z\Delta \tau$.
Results indicate the existence of the phase transition at $\tilde{J}_z\simeq 8.5$.
System size $L=16$ and $\tilde{J}_{xy}\equiv J_{xy}\Delta \tau=0.5$.
}\vspace{-0.5cm}
\label{fig:EC1}
\end{center}
\end{figure}

We first consider the case of the same mass of the $a$ and $b$-atoms,
$t=t_a=t_b=30/\Delta \tau$.
To investigate the phase structure, we calculate the internal energy $E_{\rm MC}$ and
specific heat $C_{\rm MC}$, which are defined as
\begin{eqnarray}
E_{\rm MC}&=&\langle (A_{\rm L}+A_z) \rangle/L^3, \nonumber  \\
C_{\rm MC}&=&\langle ((A_{\rm L}+A_z)-E_{\rm MC})^2 \rangle/L^3,
\label{EC}
\end{eqnarray}
where $L$ is the linear size of the 3D cubic lattice, and
we employ the periodic boundary condition.
To identify various phases, we also calculate the following
pseudo-spin and boson correlation functions:
\begin{eqnarray}
&& G_{\rm S}(r)={1 \over L^3}\sum_{r_0} \langle 
e^{i\Omega_{1,r_0}}e^{-i\Omega_{1,r_0+r}}\rangle, \nonumber \\
&& G_{a}(r)={1 \over L^3}\sum_{r_0} \langle 
e^{i\Omega_{2,r_0}}e^{-i\Omega_{2,r_0+r}}\rangle, \label{CF1} \\
&&G_{b}(r)={1 \over L^3}\sum_{r_0} \langle 
e^{i\Omega_{3,r_0}}e^{-i\Omega_{3,r_0+r}}\rangle,  \nonumber
\end{eqnarray}
where sites $r_0$ and $r_0+r$ are located in the same spatial 2D lattice,
i.e., Eqs.(\ref{CF1}) are the equal-time correlators.
From Eq.(\ref{CF1}), 
$G_{\rm S}(r)$ measures the FM spin order, whereas $G_{a}(r)$ and $G_{b}(r)$
measure the SF (BEC)  density of each atom.

The calculations of $E_{\rm MC}$ and $C_{\rm MC}$ shown in Fig.\ref{fig:EC1} 
clearly indicate the existence of a phase transition at $\tilde{J}_z\simeq 8.5$.
The order of the phase transition can be verified by calculating the 
density of states $N(E)$ that is defined as
\begin{equation}
[Z_{qXYZ}]=\int dE N(E) \ e^{-E}.
\label{N(E)}
\end{equation}
We found that $N(E)$ has a single-peak shape for all values of $\tilde{J}_z$
close to the phase boundary.
This observation indicates that the phase transition at $\tilde{J}_z\simeq 8.5$
is of second order.

\begin{figure}[h]
\begin{center}
\includegraphics[width=6cm]{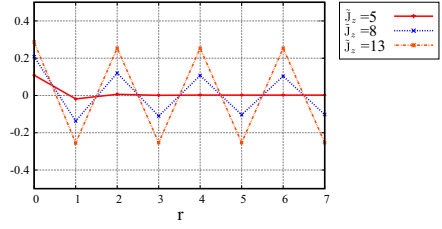}
\vspace{-0.3cm}
\caption{(Color online)
Correlation function of density difference $\Delta\rho_r$.
With the results of the correlation functions in Fig.\ref{fig:CF1},
it is concluded that there exist three phases.
}\vspace{-0.5cm}
\label{fig:Density1}
\end{center}
\end{figure}
\begin{figure}[h]
\begin{center}
\includegraphics[width=3.5cm]{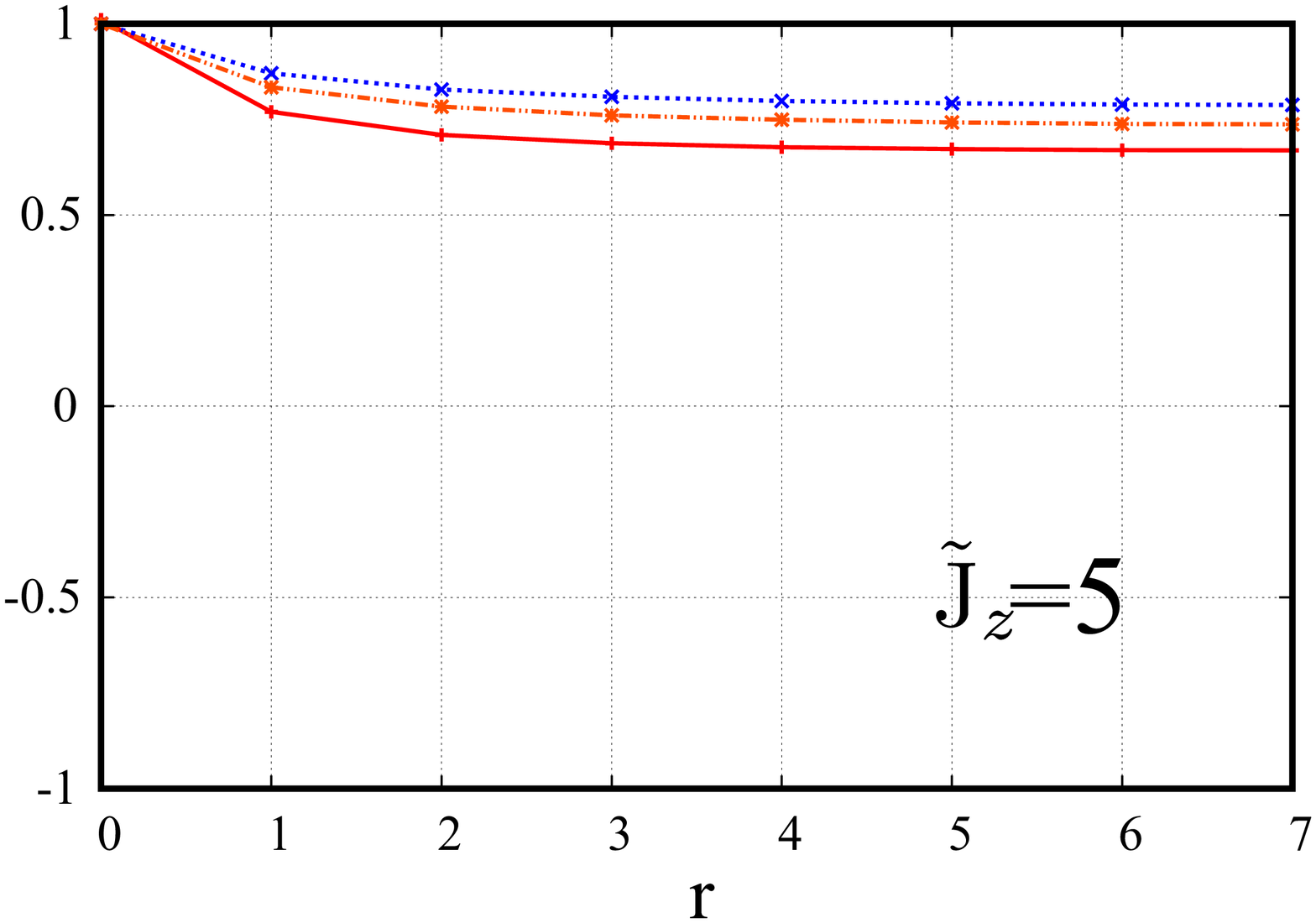}
\includegraphics[width=3.5cm]{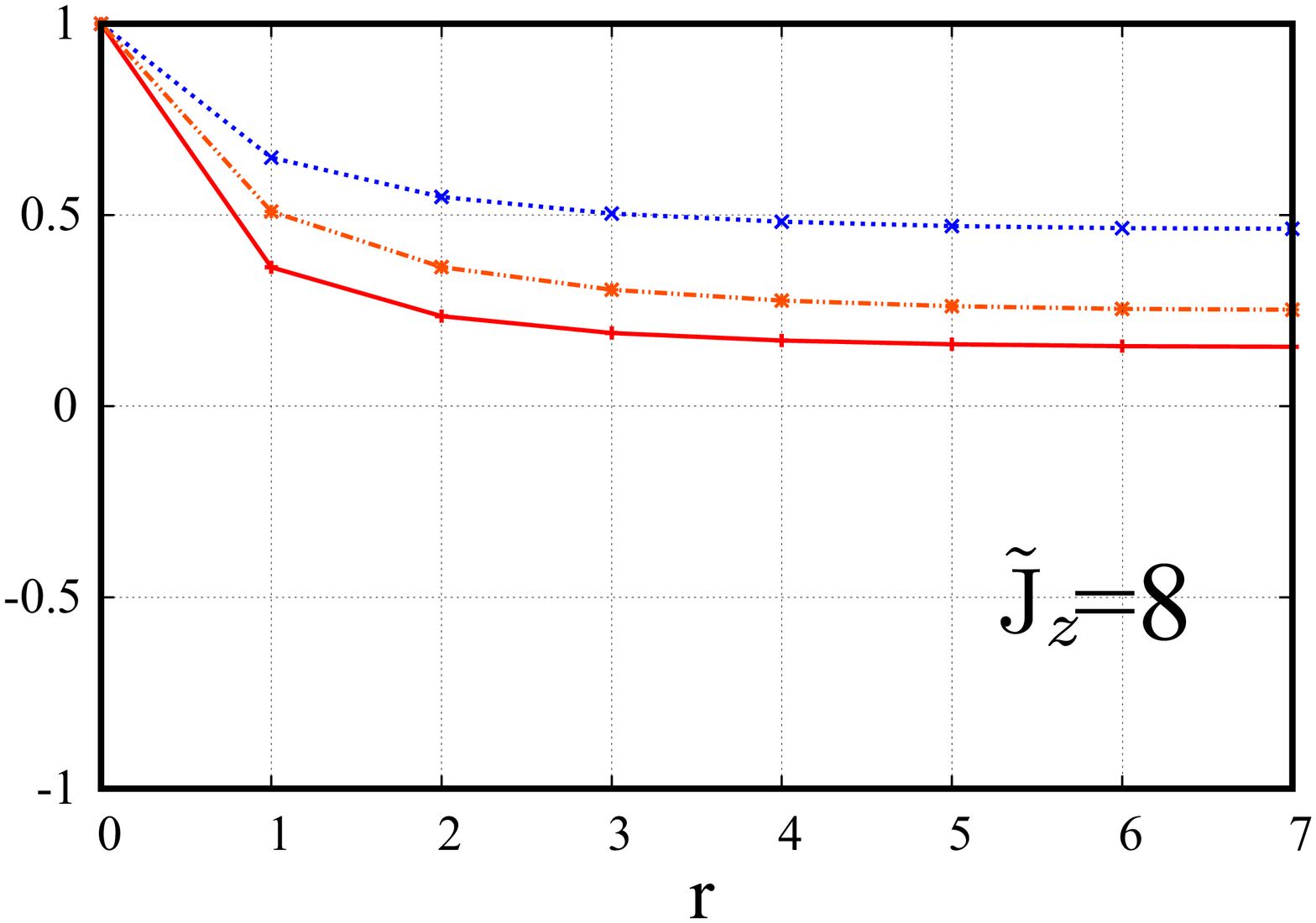}
\includegraphics[width=5.2cm]{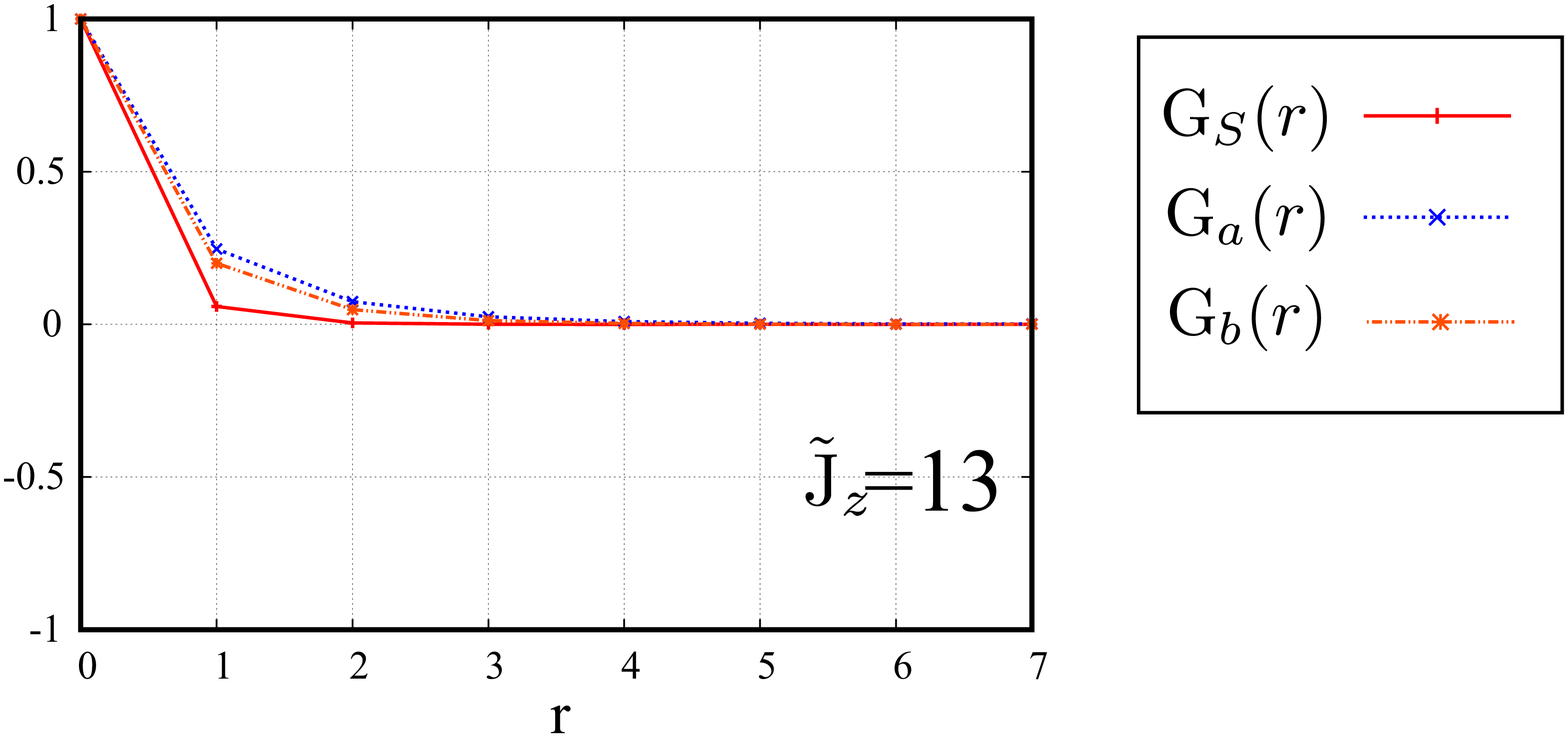}
\vspace{-0.3cm}
\caption{(Color online)
Correlation functions of boson operators defined by Eq.(\ref{CF1}).
$\tilde{J}_z=J_z\Delta \tau$.
}\vspace{-0.5cm}
\label{fig:CF1}
\end{center}
\end{figure}
\begin{figure}[h]
\begin{center} \vspace{0.5cm}
\includegraphics[width=7cm]{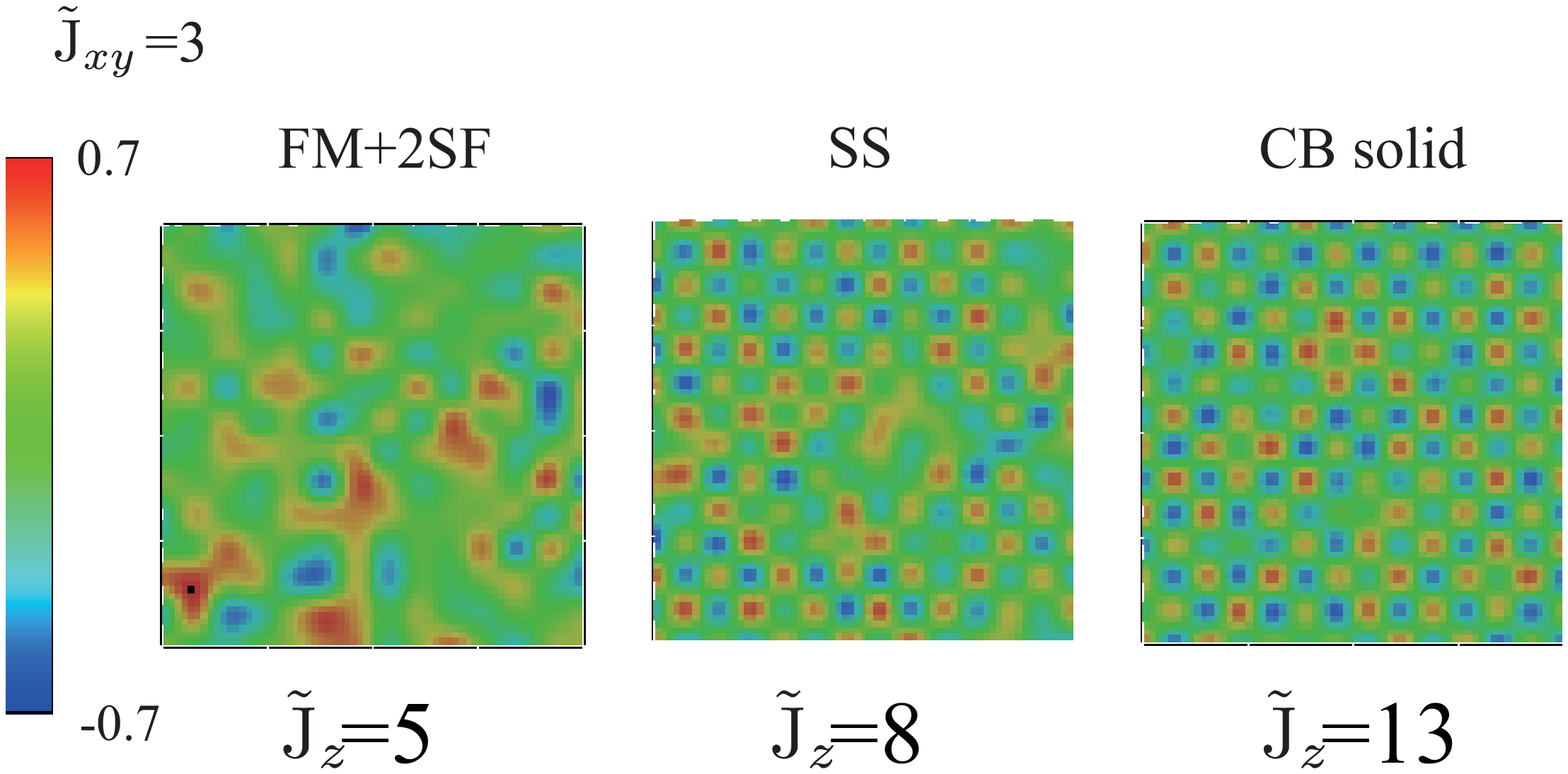} \vspace{0.5cm} \\
\includegraphics[width=7.3cm]{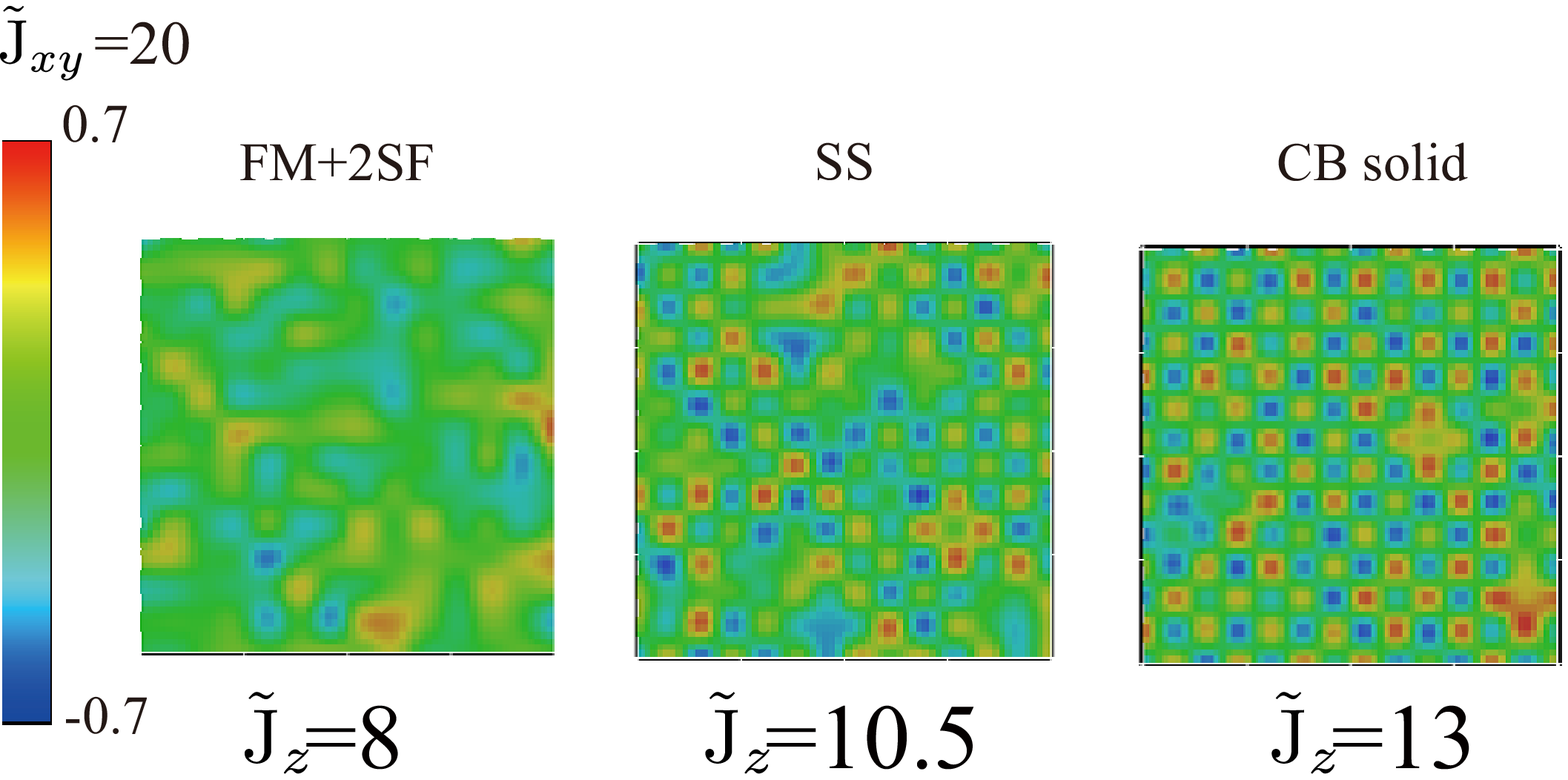}
\vspace{-0.3cm}
\caption{(Color online)
Snapshots of density difference $\Delta \rho_r$ in various phases for 
$\tilde{J}_{xy}\equiv J_{xy}\Delta \tau=3$ and $20$.
The FM+2SF state has a much more homogeneous
density for $\tilde{J}_{xy}=20$ than for $\tilde{J}_{xy}=3$.
}\vspace{-0.5cm}
\label{fig:Density_snap}
\end{center}
\end{figure}

We also measured the correlation functions in Eq.(\ref{CF1}) for various 
$\tilde{J}_z$ values, and show the results in Figs.\ref{fig:Density1} and
\ref{fig:CF1}.
It is clear that, for $\tilde{J}_z=5$, the homogeneous superfluid forms.
For $\tilde{J}_z=13$, the CB solid state without the SF LRO appears.
However, at $\tilde{J}_z=8$, the measured correlation functions
clearly indicate the existence of the SS, i.e., the density correlation
function exhibits the CB diagonal order and the boson correlations also
have the SF order. 
Snapshots of the density difference $\Delta \rho_r$
in various phases are also shown in Fig.\ref{fig:Density_snap}.
As the specific heat $C_{\rm MC}$ exhibits no anomalous behavior besides
the one at $\tilde{J}_z\simeq 8.5$, which is the phase transition from the SS 
to the solid, the transition from the SF to the SS must be of higher order or 
a crossover.
This result is in good agreement with the observation using the GP theory
given in the previous subsection.

\subsection{Monte-Carlo simulations: Case of different masses}

\begin{figure}[h]
\begin{center} 
\includegraphics[width=7cm]{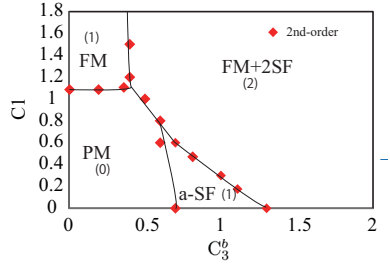} 
\vspace{-0.3cm}
\caption{(Color online)
Phase diagram of bosonic t-J model  with mass difference 
$t_a=2t_b$ and $\tilde{J}_z\ll 1$.
$C^b_3\equiv t_b\rho_3(1-\rho_3)\Delta \tau/4$ and $\rho_3=0.3$.
The numbers in parentheses denote the number of Nambu-Goldstone bosons.
This phase diagram was obtained in Ref.8.
}\vspace{-0.5cm}
\label{fig:PD_mass_diff}
\end{center}
\end{figure}
\begin{figure}[h]
\begin{center} 
\includegraphics[width=6cm]{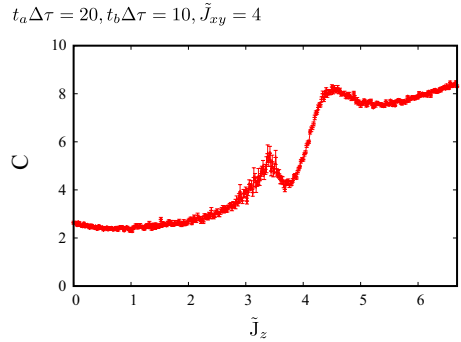} 
\vspace{-0.3cm}
\caption{(Color online)
Specific heat $C$ as a function of $\tilde{J}_z$ for the system with mass difference. 
Other parameters are $t_a\Delta \tau=20, t_b\Delta \tau=10$, and $\tilde{J}_{xy}=4$.
}\vspace{-0.5cm}
\label{fig:C_mass_diff}
\end{center}
\end{figure}
\begin{figure}[h]
\begin{center} 
\includegraphics[width=5cm]{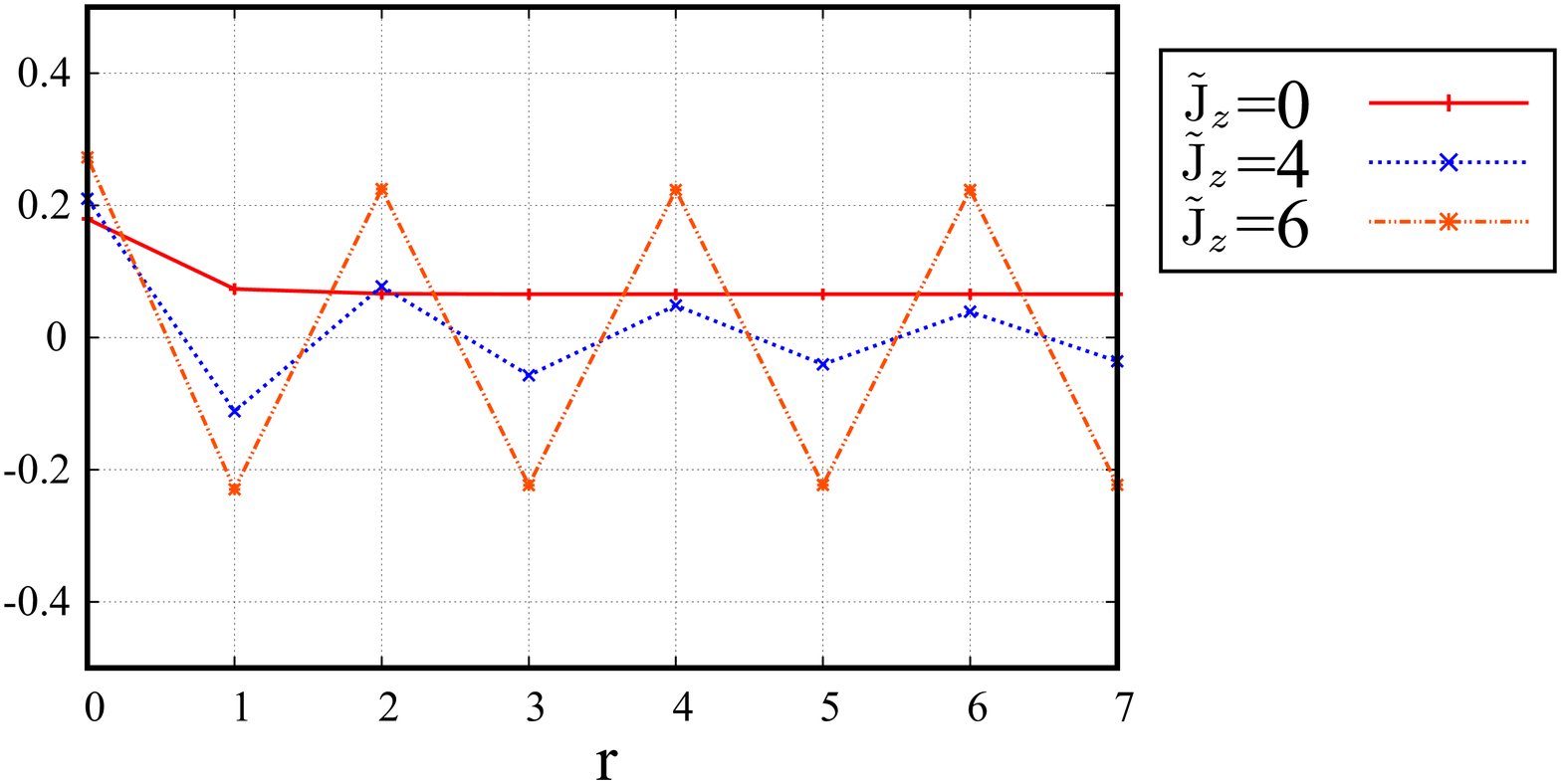} \\
\includegraphics[width=7cm]{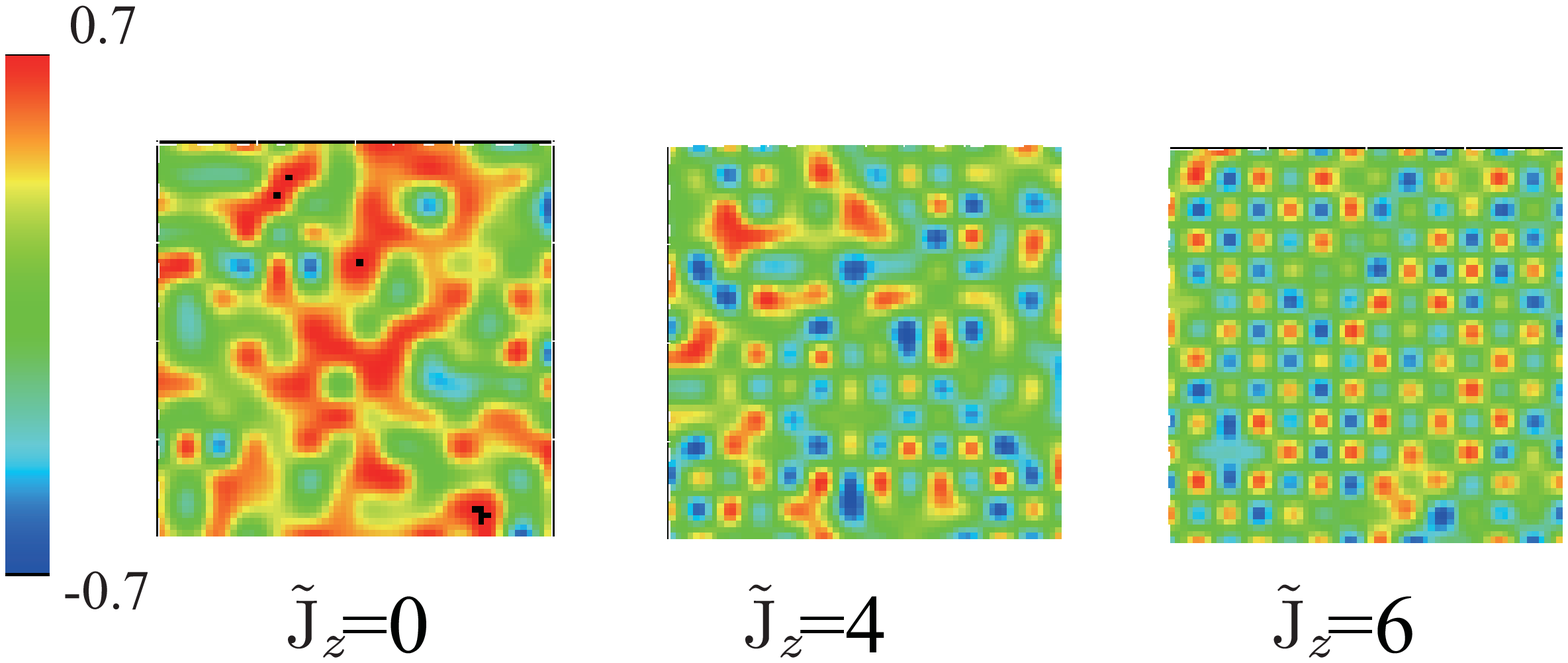} \vspace{-0.5cm}
\caption{(Color online)
Density correlation and density snapshot for SF of $a$ atom in the system
with mass difference. 
}
\label{fig:density_mass_diff}
\end{center}
\end{figure}
\begin{figure}[h]
\begin{center} 
\includegraphics[width=3.5cm]{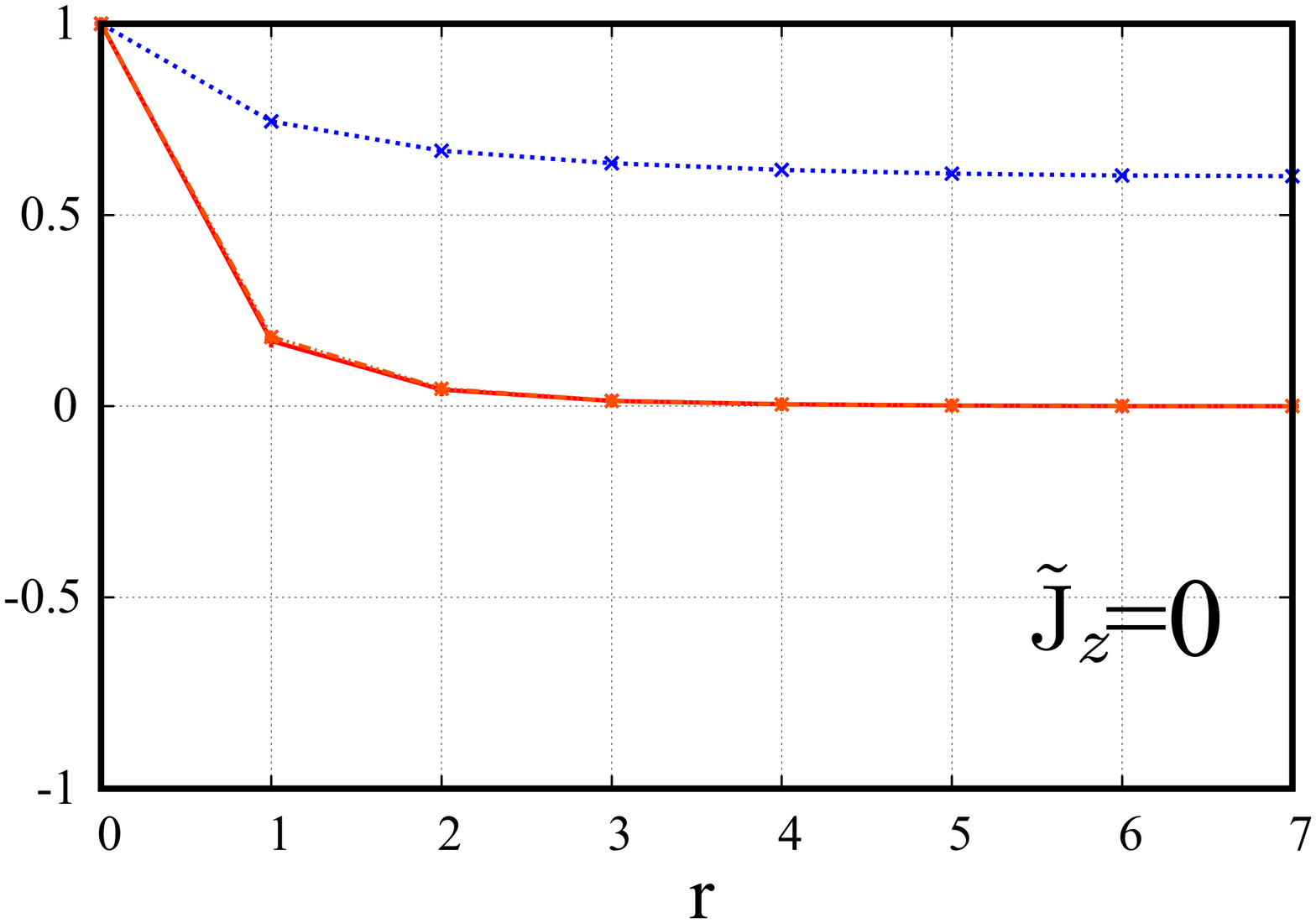} 
\includegraphics[width=3.5cm]{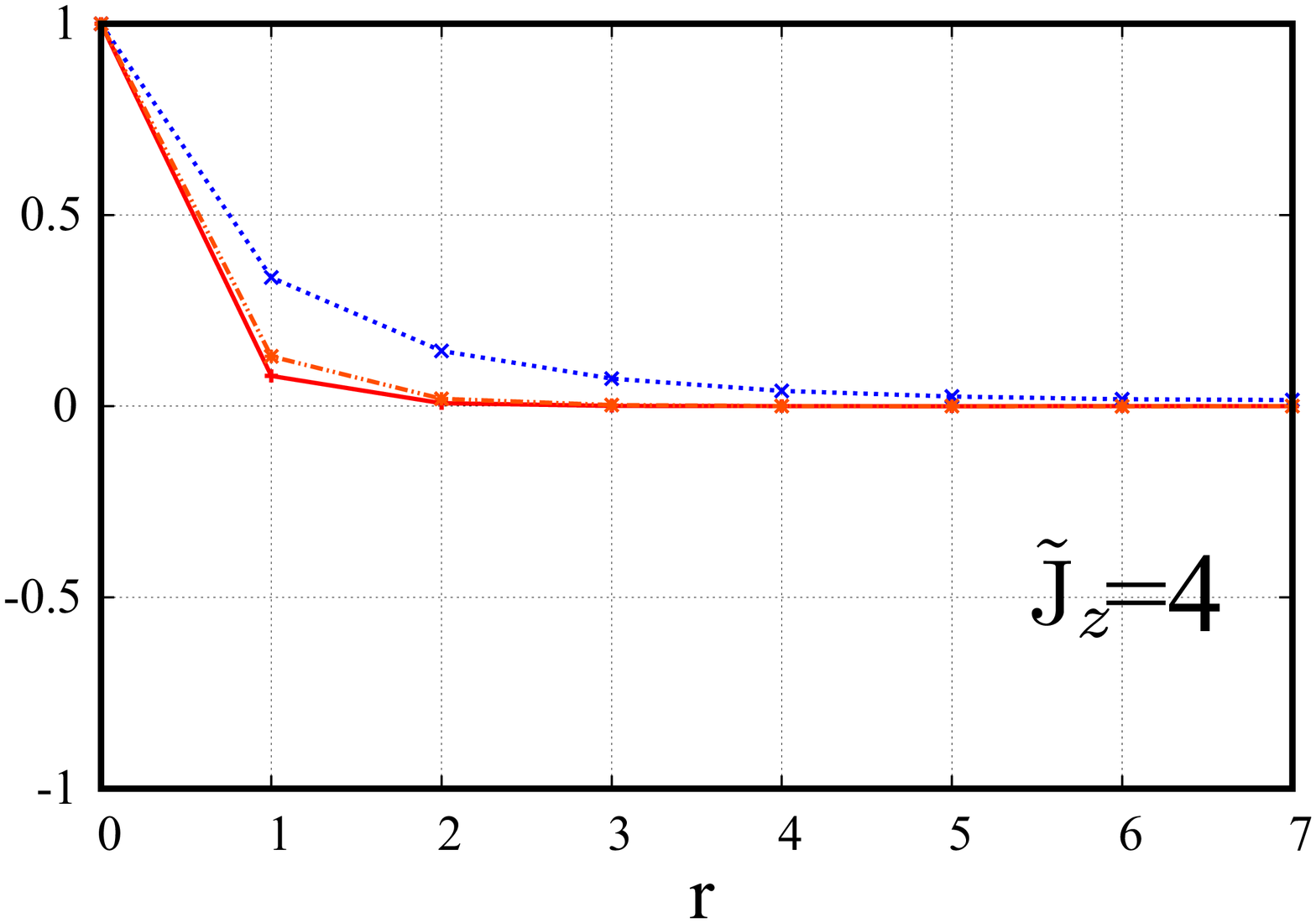} 
\includegraphics[width=5cm]{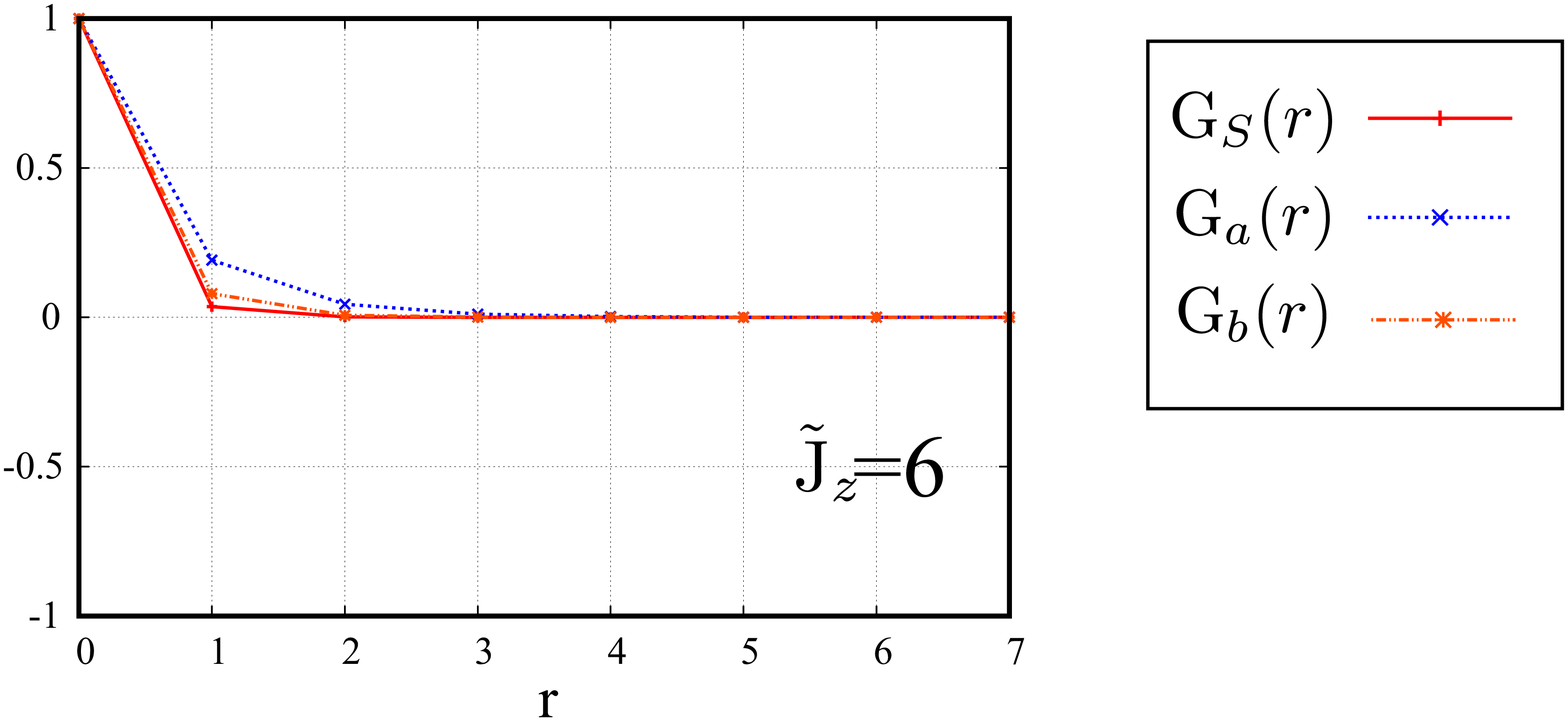} 
\vspace{-0.3cm}
\caption{(Color online)
Various correlation functions as a function of $\tilde{J}_z$ in the system
with mass difference. 
}\vspace{-0.5cm}
\label{fig:crr_mass_diff}
\end{center}
\end{figure}

Let us turn to the case of the mass difference such as $t_a=2t_b$.
The total filling factor of atoms is 0.7 as in the previous subsections.
The phase diagram of that system at $J_z\simeq 0$ was obtained in the 
previous study\cite{KSI}.
See Fig.\ref{fig:PD_mass_diff}.
In contrast to to the same-mass case, there exists a SF state of the $a$-atom
without BEC of the $b$-atom. (We call it $a$-SF in Fig.\ref{fig:PD_mass_diff}.)
We are particularly interested in how the SF phase of
$a$-atom evolves as $\tilde{J}_z$ is increased.
On the other hand
for the phase of FM+2SF with $\tilde{J}_z=0$, we have verified that there appears 
the SS for intermediate values of $\tilde{J}_z$ and the CB solid for large $\tilde{J}_z$, 
as $\tilde{J}_z$ is increased as in the same-mass case.

In Fig.\ref{fig:C_mass_diff}, we show the specific heat $C$ as a function of 
$\tilde{J}_z$ for the system with the masses $t_a=2t_b=20/\Delta\tau$
and $\tilde{J}_{xy}=4$.
At $\tilde{J}_z=0$, the system is in the single SF of $a$-atom.
As $\tilde{J}_z$ is increased,
there appear a sharp peak at $\tilde{J}_z \simeq 3.3$ and a jump at $\tilde{J}_z \simeq 4.2$.
In Fig.\ref{fig:density_mass_diff}, we show the density correlation function
and also snapshot of the density.
From the results in Fig.\ref{fig:density_mass_diff}, the states for $\tilde{J}_z=0$ 
and $\tilde{J}_z=6$ have a homogeneous and CB density, respectively, 
whereas for $\tilde{J}_z$=4, only a short-range order of the CB type exists.
Furthermore, the correlation functions in Fig.\ref{fig:crr_mass_diff} show that
the state for $\tilde{J}_z=4$ does not have the SF order, and therefore, it is not a SS.
Therefore, the peak at $\tilde{J}_z \simeq 3.3$ in $C$ corresponds to the phase transition
from the $a$-atom SF to the disordered state without any LRO's, whereas
the jump at $\tilde{J}_z \simeq 4.2$ corresponds to the transition from the 
disordered state to the CB solid.
The appearance of the disordered state without any LRO's at intermediate 
$\tilde{J}_z$ stems from the competition of the hopping terms and 
the pseudospin AF coupling.
Note that
the nonexistence of the SS in the present case is consistent with the phase
diagram of a single-boson system in the square lattice, in which the SS does
not form\cite{SSsquare}.

\subsection{Finite-temperature phase transition of the SS and CB solid}

In this subsection, we study the finite-temperature phase diagram
of the SS.
In the previous research\cite{OI}, a system of the hard-core boson in a triangular
lattice was studied.
In that system, a SS forms as a result of competition of
the nearest-neighbor repulsion and the hopping amplitude.
As the temperature is increased, two phase transitions take place,
i.e., at the first transition, the SF is lost, and at the second one,
the solid order is lost.
This result was obtained by finite-temperature MC simulations.

\begin{figure}[h]
\begin{center} 
\includegraphics[width=6cm]{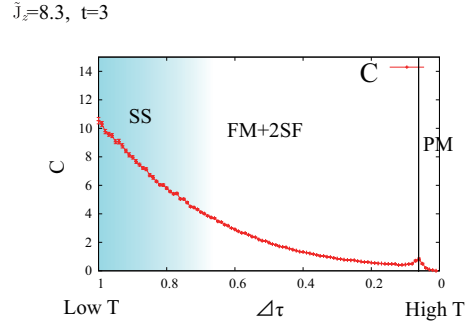} 
\vspace{-0.3cm}
\caption{(Color online)
Finite temperature phase diagram.
$\tilde{J}_z=J_z\Delta \tau$ and $T\propto 1/\Delta \tau$.
PM (paramagnetic) state stands for the disordered state without any orders.
}\vspace{-0.5cm}
\label{fig:finiteT}
\end{center}
\end{figure}

In this subsection, we investigate how the SS, which exists at vanishing
(or at very low) temperature, evolves as system temperature is increased.
This study is performed by using quantum MC simulations. 
Temperature of the system $T$ is related to the system size of
the imaginary-time direction $L_\tau$ and the time slice $\Delta \tau$ as
$k_{\rm B}T={1 \over L_\tau\Delta \tau}$.
For the MC simulation, we take $L_\tau$ as large as $L_s$ for the linear size
of the spatial direction, i.e., $L_\tau=L_s=L$.
Then as $\Delta \tau$ is decreased, $T$ becomes higher.
As Eq.(\ref{C123}) shows, the coefficients of the action $A_{qXYZ}$ vary
with $\Delta \tau$,  and the system tends to be quasi-one-dimensional for a
small $\Delta \tau$.
That is, $J_{xy}\Delta \tau, \ t\Delta \tau\rightarrow \mbox{small}$,
whereas ${1 \over V_0\Delta \tau}\rightarrow \mbox{large}$.
This behavior of the coefficients denotes that the three-dimensional system 
at low $T$ tends to be one-dimensional for increasing $T$.
Therefore there exists a phase transition from an ordered state to a disordered
state as $T$ is increased, as is naturally expected.
In particular, it is interesting to see how the two orders of the SS, i.e., 
the SF and CB solid, disappear as $T$ is increased.
More precisely concerning the SF order, the genuine ordered state
turns to the state with {\em quasi-LRO} as the system is quantum 2D
at a finite temperature.
Therefore, it is expected that a phase transition from the SF to 
the disordered state belongs to the universality class of the Kosterlitz-Thouless (KT)
transition.

We investigated the above problem by the present quantum MC simulations. 
We consider the same-mass case $t_a=t_b=t$ and the total filling factor 0.7, 
and focus on the SS observed in Sec. 3.2.
The obtained phase diagram is shown in Fig.\ref{fig:finiteT}.
The SS first loses the CB solid order and becomes the FM+2SF state at 
an intermediate $T$ and then loses the SF order for a higher $T$.
However as the calculation of the specific heat $C$ in Fig.\ref{fig:finiteT} indicates,
the transition from the SS to FM+2SF has no sharp phase boundary
as in the case of the quantum phase transition induced by varying $J_z$.
On the other hand, the transition at $\Delta\tau \simeq 0.06$ exhibits
a small peak in $C$.
This transition is expected to be in the universality class of the KT transition
as we explained above.
In fact, we verified that the boson correlation functions in the FM+2SF
exhibit the quasi-LRO instead of the genuine LRO, whereas those in the
PM state decay very rapidly as a function of the distance $r$.

\begin{figure}[h]
\begin{center} 
\includegraphics[width=6cm]{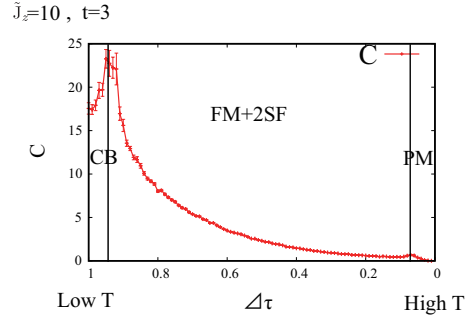} 
\vspace{-0.3cm}
\caption{(Color online)
Finite temperature phase diagram of CB solid.
At intermediate temperature, the 2SF state appears.
$\tilde{J}_z=J_z\Delta \tau$ and $T\propto 1/\Delta \tau$.
}\vspace{-0.5cm}
\label{fig:finiteT_CBS}
\end{center}
\end{figure}

Finally, let us turn to the finite-temperature phase diagram of the CB solid.
We consider the case of the equal-hopping amplitude again, $t_a=t_b=t$.
In Fig.\ref{fig:finiteT_CBS}, we show the finite-temperature phase
diagram obtained by the MC simulations.
As $T$ is increased, the solid loses the CB order and the SF appears simultaneously.
Because of the thermal fluctuations, the effect of the $J_z$-term is weakened
and then a homogeneous state with the SF is realized at an intermediate $T$.
Calculation of the specific heat $C$ in Fig.\ref{fig:finiteT_CBS} indicates 
that this phase transition is of second order, although a first-order phase 
transition is expected as this is a phase transition from the state with 
the CB order to the state with the SF.
The fact that the FM+2SF state has only the quasi-LROs, not the genuine LRO's,
probably weakens the phase transition from first to second order.
As the temperature increases further, the thermal  fluctuations destroy 
the SF and the state without any orders appears.


\section{Conclusion}

In this paper, we have studied the two-component cold Bose gas in a square optical
lattice.
In the strong on-site repulsion case, the system is described by the bosonic t-J model.
Using the GP equations and MC simulations, we investigated the phase diagram
of the model.
In particular, we are interested in how the SF evolves as the coefficient of the
$J_z$-term of the pseudospin interaction is increased. 
Both the GP theory and the quantum MC simulations show that the SS
forms at some critical value of $J_z$, and then the phase transition from the SS to
the CB solid takes place at the second critical value of $J_z$.
This result is consistent with the phase diagram obtained in a previous 
study\cite{KSI}.
In the present study, we have not assumed any density pattern for the boson
densities in contrast to our previous analysis.

In the GP theory, we have proposed a method of identifying phase boundaries 
by calculating the internal energy and ``specific heat".
In the MC simulations, we treated the boson densities at each site as variational
parameters.
The results obtained by the above two methods are in good agreement.

Finally, we studied the finite-temperature phase diagrams of the SS and CB solid,
and obtained interesting results, such as, the order of the phase transitions
and the appearance of the SF order from the CB solid as the temperature 
increases.

\bigskip 
\acknowledgments 
This work was partially supported by Grant-in-Aid
for Scientific Research from the Japan Society for the 
Promotion of Science under Grant No. 23540301.



\begin{thebibliography}{9}
\bibitem{initial}
As initial states for the numerical evolution of the GP equations,
fairly homogeneous states with small fluctuations in
amplitude and phase generated by random numbers were prepared.
\bibitem{FN_EGP}We numerically verified that inclusion of the optical
lattice term into the energy does not change its qualitative behavior.
\end{thebibliography}
\end{document}